\documentclass[longauth]{aa} 

\usepackage{natbib}
\usepackage{graphicx, color, booktabs}
\usepackage{cuted}
\usepackage{gensymb}
\usepackage{supertabular}
\usepackage{threeparttable}
\usepackage{listings}
\usepackage{xcolor}
\usepackage{textcomp}
\usepackage{pdflscape}
\usepackage{caption}
\usepackage{orcidlink}
\usepackage{dblfloatfix}

\usepackage{txfonts}
\usepackage{longtable}
\usepackage{comment}
\usepackage{hyperref}
\usepackage{multicol, multirow, colortbl}
\usepackage[export]{adjustbox}

\def\MCN{\mbox{CH$_3$CN}}

\def\g31{G31}

\begin{document}

   \title{ALMAGAL}
   \subtitle{IX. The chemical complexity of AG318.9477$-$00.1960:\\A line-identification template for ALMAGAL}
   \titlerunning{ALMAGAL\,X: The chemical complexity of AG318.9477$-$00.1960}

   \author{J. Allande \inst{1,\ref{Arcetri}}\orcidlink{0009-0007-4060-0560}, 
          M.\ T.\ Beltr\'an \inst{\ref{Arcetri}}\orcidlink{0000-0003-3315-5626}, 
          V.\ M.\ Rivilla \inst{\ref{CAB}}\orcidlink{0000-0002-2887-5859}, 
          \'A. L\'opez-Gallifa \inst{\ref{CAB}}\orcidlink{0000-0001-6049-9366}, 
          C. Y. Law \inst{\ref{Arcetri}}\orcidlink{0000-0003-1964-970X},
          \'A.\ S\'anchez-Monge \inst{\ref{ICE},\ref{IEEC}}\orcidlink{0000-0002-3078-9482},
          C.\ Battersby \inst{\ref{CONNECTICUT}},
          M.\ Benedettini \inst{\ref{ROMA}}\orcidlink{0000-0002-3597-7263},
          H.\ Beuther \inst{\ref{mpia}}\orcidlink{0000-0002-1700-090X},
          C.\ L.\ Brogan \inst{\ref{NRAO}}\orcidlink{0000-0002-6558-7653},
          L.\ Bronfman \inst{\ref{CHILE}}\orcidlink{0000-0002-9574-8454},
          S.\ D.\ Clarke \inst{\ref{TAIWAN}}\orcidlink{0000-0001-9751-4603},
          L.\ Colzi \inst{\ref{CAB}}\orcidlink{0000-0001-8064-6394},
          D.\ Elia \inst{\ref{ROMA}}, 
          F.\ Fontani \inst{\ref{Arcetri}},
          G.\ A.\ Fuller \inst{\ref{MANCHESTER},\ref{GERM}}\orcidlink{0000-0001-8509-1818},
          T.\ R.\ Hunter \inst{\ref{NRAO}}\orcidlink{0000-0001-6492-0090},
          P.\ T.\ P.\ Ho \inst{\ref{TAIWAN2},\ref{EAO}}\orcidlink{0000-0002-3412-4306},
          K.\ G.\ Johnston \inst{\ref{INB}}\orcidlink{0000-0003-4509-1180},
          B.\ M.\ Jones \inst{\ref{GERM}}\orcidlink{0000-0002-0675-0078},
          K.\ -T.\ Kim \inst{\ref{KOREA_ASSI},\ref{UST_KOREA}}\orcidlink{0000-0003-2412-7092},
          P.\ D. \ Klaassen \inst{\ref{UKATC}}\orcidlink{0000-0001-9443-0463},
          R.\ S.\ Klessen\inst{\ref{ITA},\ref{IWR}}\orcidlink{0000-0002-0560-3172}, 
          R.\ Kuiper \inst{\ref{DUISBURG}}\orcidlink{0000-0003-2309-8963},
          D.\ C.\ Lis \inst{\ref{JPL}}, 
          C.\ Mininni \inst{\ref{ROMA}}\orcidlink{0000-0002-2974-4703},
          S.\ Molinari \inst{\ref{ROMA}}\orcidlink{0000-0002-9826-7525},
          T.\ M\"{o}ller \inst{\ref{GERM}}\orcidlink{0000-0002-9277-8025},
          L.\ Moscadelli \inst{\ref{Arcetri}},
          A.\ Nucara \inst{\ref{ROMA},\ref{ROMA_UNI}}\orcidlink{0009-0005-9192-5491},
          S.\ Pezzuto \inst{\ref{ROMA}}\orcidlink{0000-0001-7852-1971}, 
          P.\ Schilke \inst{\ref{GERM}}\orcidlink{0000-0003-2141-5689},
          E.\ Schisano \inst{\ref{ROMA}}\orcidlink{0000-0003-1560-3958},
          F.\ van der Tak \inst{\ref{SRON},\ref{GRONINGEN}}\orcidlink{0000-0002-8942-1594},
          Y.\ Tang \inst{\ref{TAIWAN2}}\orcidlink{0000-0002-0675-276X},
          A.\ Traficante \inst{\ref{ROMA}}, 
          S.\ Walch \inst{\ref{GERM},\ref{COLOGNE}}\orcidlink{0000-0001-6941-7638},
          Q.\ Zhang \inst{\ref{HARVARD}}\orcidlink{0000-0003-2384-6589},
          T.\ Zhang \inst{\ref{GERM}, \ref{CHINA}}\orcidlink{0000-0002-1466-3484}
          }
    \authorrunning{J.\ Allande et al.}
    \institute{Universit\`a Degli Studi di Firenze, Via G. Sansone 1, 50019 Sesto Fiorentino, Firenze, Italy
              \
              \email{jofre.allandegonzalez@unifi.it}
         \and
            \label{Arcetri} INAF-Osservatorio Astrofisico di Arcetri, Largo E.\ Fermi 5, 50125 Firenze, Italy 
        \and
            \label{CAB} Centro de Astrobiolog\'ia (CAB), CSIC-INTA, Ctra. de Ajalvir, km. 4, Torrej\'on de Ardoz, E-28850 Madrid, Spain
        \and
            \label{ICE}
            Institut de Ci\`encies de l'Espai (ICE), CSIC, Campus UAB, Carrer de Can Magrans s/n, E-08193, Bellaterra, Barcelona, Spain
        \and
            \label{IEEC}
            Institut d'Estudis Espacials de Catalunya (IEEC), E-08860, Castelldefels, Barcelona, Spain
        \and
            \label{CONNECTICUT}
            University of Connecticut, Department of Physics, 196A Auditorium Road Unit 3046, Storrs, CT 06269 USA
        \and
            \label{ROMA}
            INAF-Istituto di Astrofisica e Planetologia Spaziale, Via Fosso del Cavaliere 100, 00133 Roma, Italy
        \and
            \label{mpia}
            Max Planck Institute for Astronomy, K\"onigstuhl 17, 69117 Heidelberg, Germany
        \and
            \label{NRAO}
            National Radio Astronomy Observatory, 520 Edgemont Rd, Charlottesville VA 22903 USA
        \and
            \label{CHILE}
            Departamento de Astronomía, Universidad de Chile, Casilla 36-D, Santiago, Chile
        \and
            \label{TAIWAN}
            Department of Physics, National Cheng Kung University, No.1, University Road, Tainan City 70101, Taiwan
        \and
            \label{MANCHESTER}
            Jodrell Bank Centre for Astrophysics, Oxford Road, The University
            of Manchester, Manchester M13 9PL, UK
        \and
            \label{GERM}
            I. Physikalisches Institut, Universit{\"a}t zu K{\"o}ln, Z{\"u}lpicher Stra{\ss}e 77, 50937 K{\"o}ln, Germany
        \and
            \label{TAIWAN2}
            Institute of Astronomy and Astrophysics, Academia Sinica, 11F of ASMAB, AS/NTU No. 1, Sec. 4, Roosevelt Road, Taipei 10617, Taiwan
        \and
            \label{EAO}
            East Asian Observatory, 660 N. A’ohoku, Hilo, Hawaii, HI 96720, USA
        \and
            \label{INB}
            School of Engineering and Physical Sciences, Isaac Newton Building, University of Lincoln, Brayford Pool, Lincoln LN6 7TS, UK
        \and
            \label{KOREA_ASSI}
            Korea Astronomy and Space Science Institute, 776 Daedeokdae-ro, Yuseong-gu, Daejeon 34055, Republic of Korea
        \and
            \label{UST_KOREA}
            University of Science and Technology, Korea (UST), 217 Gajeong-ro, Yuseong-gu, Daejeon 34113, Republic of Korea
        \and
            \label{UKATC}
            UK Astronomy Technology Centre, Royal Observatory Edinburgh, Blackford Hill, Edinburgh EH9 3HJ, UK
        \and
            \label{ITA} Universit\"{a}t Heidelberg, Zentrum f\"{u}r Astronomie, Institut f\"{u}r Theoretische Astrophysik, Albert-Ueberle-Str.\ 2, 69120 Heidelberg, Germany
        \and        
            \label{IWR} Universit\"{a}t Heidelberg, Interdisziplin\"{a}res Zentrum f\"{u}r Wissenschaftliches Rechnen, Im Neuenheimer Feld 225, 69120 Heidelberg, Germany
        \and
            \label{DUISBURG}
            Faculty of Physics, University of Duisburg-Essen, Lotharstra{\ss}e 1, D-47057 Duisburg, Germany
        \and
            \label{JPL}
            Jet Propulsion Laboratory, California Institute of Technology, 4800 Oak Grove Drive, Pasadena, CA 91109, USA
        \and
            \label{ROMA_UNI}
            Dipartimento di Fisica, Università di Roma Tor Vergata, Via della Ricerca Scientifica 1, I-00133 Roma, Italy
        \and
            \label{SRON}
            SRON Netherlands Institute for Space Research, Landleven 12, 9747, AD Groningen, The Netherlands
        \and
            \label{GRONINGEN}
            Kapteyn Astronomical Institute, University of Groningen, 9700, AV Groningen, The Netherlands
        \and
            \label{COLOGNE}
            Center for Data and Simulation Science, University of Cologne, Germany
        \and
            \label{HARVARD}        
            Center for Astrophysics $\vert$ Harvard \& Smithsonian, 60 Garden Street, Cambridge, MA, 02138, USA
        \and
            \label{CHINA}
            Research Center for Computational Earth and Space Science, Zhejiang Laboratory, Hangzhou, China
        }


 
  \abstract
   {High-mass star-forming regions are rich in complex organic molecules (COMs), which are carbon-bearing species with at least six atoms. Their formation pathways remain debated. The ALMA evolutionary study of high-mass protocluster formation in the GALaxy (ALMAGAL) survey provides a unique opportunity to probe this chemical complexity in a large statistical significant sample of high-mass star-forming regions.} 
   {We present a detailed molecular line analysis of one of the most chemically rich cores in the ALMAGAL sample, the high-mass core~9 in the AG318.9477$-$00.1960 clump (AG318$-$c9), located at a heliocentric distance of $\sim 10.4\,\rm kpc$. This source was selected because it combines an exceptionally high line density with a relatively simple kinematic structure, making it an optimal template for chemical analysis in the survey. We further assessed whether the emission of selected COMs, that is, ethylene glycol ($\rm (CH_2OH)_2$; EG), glycolaldehyde ($\rm CH_2(OH)CHO$; GA), and methyl formate ($\rm CH_3OCHO$; MF), can be used to trace the innermost regions of hot molecular cores (HMCs).}
   {We analysed ALMA Band~6 observations ($\sim 217-221\,$GHz). Spectral line identification and local thermodynamic equilibrium modelling were performed using the software called MAdrid Data CUBe Analysis (MADCUBA). We derived the physical parameters, including the column density ($N$), excitation temperature ($T_{\rm ex}$), velocity, line width, and molecular abundances relative to $\rm H_2$, for all detected species. The chemical inventory of AG318$-$c9 was compared with that of the HMC $\rm G31.41+0.31$ (G31). In addition, we performed a pixel-by-pixel analysis of EG, GA, and MF to generate spatially resolved $N$ and $T_{\rm ex}$ maps and corresponding radial profiles.}
   {We report the detection of 65 molecular species, including 31 main species and 34 isotopologues and vibrationally excited species. Of these, 44 are O-bearing species, 28 are N-bearing, 8 are S-bearing, and 2 are Si-bearing. While AG318$-$c9 exhibits lower abundances than G31 overall, it shows detections of species that have not yet been reported in G31, such as ethyl formate ($\rm C_2H_5OCHO$). Moreover, 12 species depart from the general trend, displaying relative overabundances in AG318$-$c9. The comparison also reveals a chemical differentiation between O- and N- bearing COMs, with O-bearing species systematically more abundant in G31. Regarding EG, GA, and MF, the latter is the most abundant ($X \sim 3\times 10^{-8}$), followed by EG ($X\sim 7\times 10^{-9}$) and GA ($X \sim 2\times 10^{-9}$). The $N$ and $T_{\rm ex}$ maps suggest that MF is the most spatially extended species, whereas EG and GA are more compact and confined to the innermost hot region.}
   {AG318$-$c9 reveals a rich chemical inventory characteristic of an HMC. The chemical comparison with G31 suggests that AG318$-$c9 is a less evolved hot core dominated by the recent sublimation of ice mantles, in contrast to the higher abundances observed in the possible more evolved G31. MF emerges as an excellent tracer of the gas temperature in HMCs owing to its wide spatial extent and wide dynamic range of the excitation temperature. In contrast, EG and GA are more compact and preferentially trace the innermost high-density regions of the core, suggesting a high sublimation temperature threshold and supporting a shared grain-surface formation pathway. While EG and MF seem to be partly limited by sensitivity, GA reflects an intrinsically central distribution.}


   \maketitle
%

\section{Introduction}
\label{introduction}
Astrochemistry is inherently a multidisciplinary field that involves physics, chemistry, biology, and geology to investigate the formation and evolution of molecules in the interstellar medium (ISM). Unveiling the chemistry in a particular environment provides crucial insight into the emergence of molecular complexity, the dominant chemical processes, and how these are inherited at different evolutionary stages. Furthermore, astrochemical studies provide constraints on the physical conditions and lifetimes of these regions \citep{reviewJorgensen}. At its core, astrochemistry studies the potential precursors of life, and together with astrobiology, addresses one of the most fundamental open questions in science: the origin of life.

As of November 2025, approximately 340 molecules have been detected in the ISM or in circumstellar shells \citep{cdms_molecules}. They include the so-called complex organic molecules (COMs), which are defined as carbon-bearing species with at least six atoms \citep{herbst}. COMs play a central role as they represent key steps towards increasing chemical complexity, potentially leading to the formation of prebiotic species. The latter are regarded as building blocks of life, as they share structural elements with biomolecules found in living organisms and are thought to play a role in prebiotic chemical processes \citep{caselli_ceccarelli,ceccarelli2022}. 

Although the formation of COMs remains under debate, two main and potentially complementary pathways have been proposed: (i) surface reactions on interstellar dust grains \citep{Simons2020, garrod2022}, and (ii) gas-phase chemical reactions \citep{balucani2015, skouteris2018}. During the early stages of star formation, cold molecular clouds are characterised by temperatures of $\rm \sim 10$ -- $20\,K$ and densities of $\rm \sim 10^3-10^4\,cm^{-3}$, which increase as the cloud evolves and collapses. With these conditions, atoms and simple species such as oxygen, nitrogen, and hydrogen accrete onto dust grains, forming icy mantles. In this scenario, grain-surface chemistry is dominated by hydrogenation reactions, leading to a chemical inventory dominated by simple species such as water ($\rm H_2O$), methane ($\rm CH_4$), or ammonia \citep[$\rm NH_3$;][]{tielens2013,boogert2015,linnartz2015}. As collapse proceeds and densities reach $\rm \sim 10^4-10^6\,cm^{-3}$, significant CO freeze-out occurs, enabling CO hydrogenation and the formation of species such as the formyl radical (HCO), formaldehyde ($\rm H_2CO$), and eventually, methanol ($\rm CH_3OH$), the simplest and most abundant COM \citep{watanabe2002,fuchs2009,santos2022}. Moreover, ultraviolet photons and cosmic rays enable the gradual build-up of molecular complexity \citep{oberg2016Photochemistry,Padovani2020}. As material accretes onto the protostar, the resulting accretion shocks significantly increase the central luminosity and raise the envelope temperature from 10\,K to $100$ -- $300$\,K. This thermal processing triggers the sublimation of icy mantles from dust grains, releasing species into the gas phase \citep{caselli_ceccarelli}. When they are in the gas phase, COMs can be detected with radio telescopes via their characteristic rotational transitions. 

The COMs are commonly observed in high-mass star-forming regions at different evolutionary stages due to high gas densities, intense heating occurring during the hot molecular core (HMC) phase, and efficient desorption processes that enhance their detectability and abundance \citep{Baek_2022, Coletta2020Evolutionary}. HMC regions have the richest chemistry in the Galaxy \citep{bisschop2007,belloche2013,rivilla2017}. In addition to tracing chemical complexity, COMs probe a wide range of physical conditions (temperature and density) and provide valuable insights into the physical properties, kinematics, and formation mechanisms of high-mass star-forming regions. In this context, heavy COMs (those with $\geq 8$ atoms) and prebiotic species are typically less abundant and more optically thin than standard HMC tracers such as CH$_3$CN, allowing them to better probe the innermost densest regions of high-mass cores \citep{rivilla2017}. 

However, it remains challenging to detect COMs because their emission is generally weaker than that of simpler and more abundant species. In chemically rich regions, spectra can become so densely populated with molecular transitions that individual lines overlap and cannot be uniquely identified, reaching the line-confusion limit, where increased sensitivity no longer enables secure detections of weaker species. This effect is particularly severe in turbulent environments. Consequently, a careful and detailed line identification is required to reliably assign molecular transitions and derive the physical properties of less abundant species. 

\begin{figure*}[h]
    \centering
    \includegraphics[width=\linewidth]{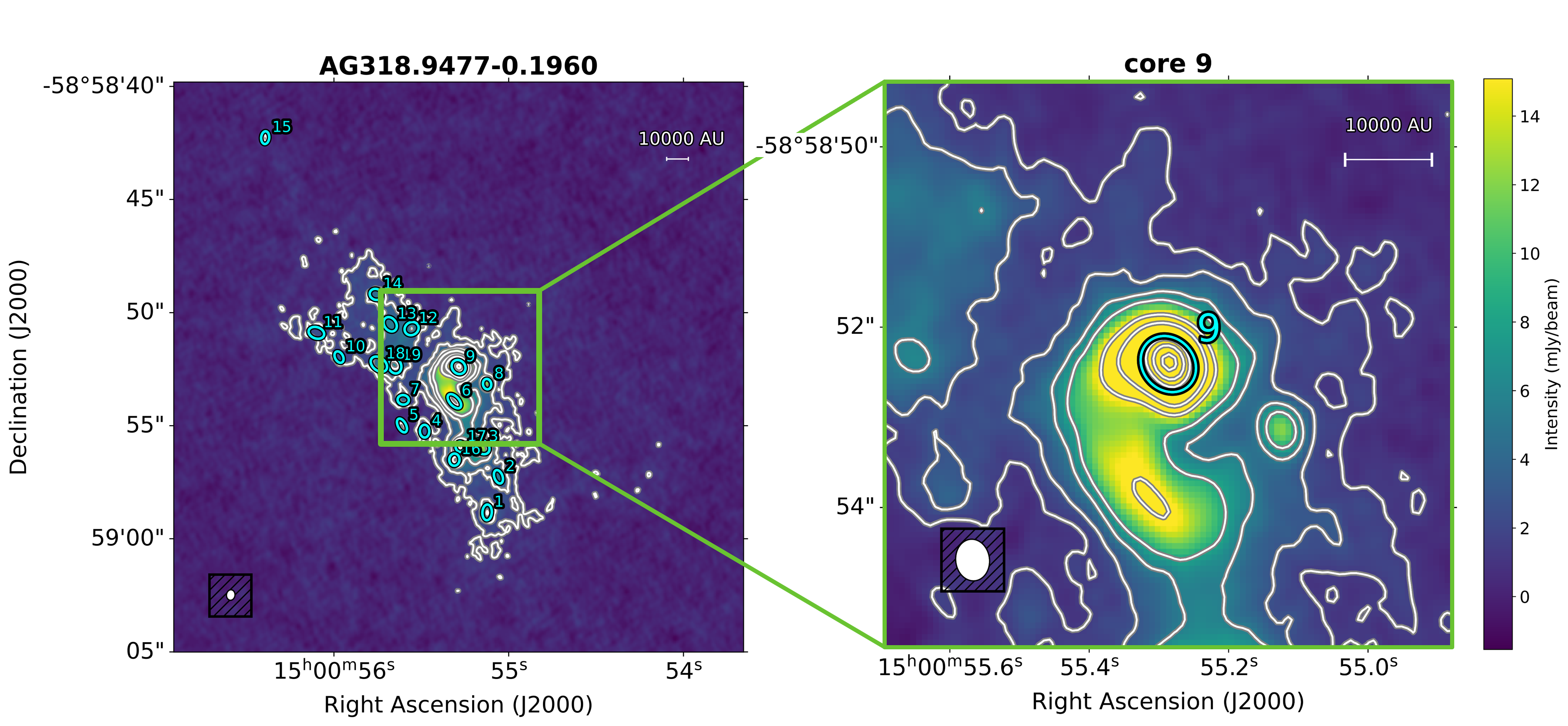}
    \caption{ALMA map of the continuum emission at 1.38\,mm of the AG318.9477$-$0.1960 clump (left) and a zoom-in on core 9 (right). The white contours correspond to $5\sigma$, $10\sigma$, $20\sigma$, $30\sigma$, $60\sigma$, $90\sigma$, $180\sigma$, $220\sigma$, and $260\sigma$, with $\sigma \sim 0.28\rm\,mJy\, beam^{-1}$. The cyan ellipses surrounded by black show the position and size of each core. In particular, the size of core~9 is $\rm 0\farcs66\times0\farcs57$, with a position angle of PA $=42\,\degree$. The synthesised beam ($\rm 0\farcs46 \times 0\farcs37$, PA$=7\,\degree$) is shown in the bottom left corner.}
    \label{AG318_core9}
\end{figure*}

Motivated by these considerations, we performed an exhaustive line identification towards core~9 of the AG318.9477$-$0.1960 clump (hereafter AG318$-$c9; Fig.~\ref{AG318_core9}), which is one of the most chemically rich cores in the ALMA evolutionary study of high-mass protocluster formation in the Galaxy (ALMAGAL) large project (2019.1.00195.L; \citealt{almagal1, almagal2}). ALMAGAL is a survey of high-mass star-forming regions in the Galactic plane that are observed with the Atacama Large Millimetre Array (hereafter ALMA; \citealt{ALMA}) in Band 6 \citep{band6}. The survey was selected from the Herschel Infrared Galactic Plane Survey (Hi-GAL, \citealt{hi-gal,hi-gal2016,elia_higal2018,hi-gal_elia2021}) and Red MSX Source survey (RMS; \citealt{hoare2005,urquhart2007,lumsden2013}) catalogues. By observing 1013 dense clumps in various evolutionary stages and Galactic environments, ALMAGAL provides a statistically significant and uniformly observed sample to investigate the physical and chemical evolution of high-mass star-forming regions. Detailed descriptions of the ALMAGAL survey characterisation \citep{almagal1}, dataset and data reduction pipeline \citep{almagal2}, fragmentation properties of the compact sources \citep{almagal3}, and early scientific results \citep{wells_2024, almagal4, almagal5, almagal6, almagal8} are available.

In a forthcoming study (Allande et al., in prep.), we will analyse the emission of three potentially chemically related COMs across the full ALMAGAL sample: GA, EG, and MF. Laboratory experiments and astrochemical models indicate that GA and EG are chemically linked and may share common formation pathways in the ISM \citep{Sorrell_2001,fedoseev2015,coutens2018,garrod2022}. MF, while not directly connected through the same synthetic routes, is the most abundant isomer of GA and provides complementary constraints on the formation of GA and EG, as all three species are expected to arise under similar physical conditions \citep{Simons2020}.

As a necessary first step towards this survey-wide analysis, we present a detailed line identification of the ALMAGAL spectral band coverage (216.98 –- 218.86\,GHz and 219.06 -- 220.94\,GHz) towards AG318$-$c9. We identify all detectable molecular species contributing to the emission in this frequency range and perform an initial analysis of the physical properties of GA, EG, and MF, with particular emphasis on assessing line blending and selecting the least affected transitions for further study. In addition, we explore the feasibility of deriving spatially resolved physical parameter maps (column density and excitation temperature) for these species, evaluating their potential as tracers of the physical conditions of dense gas \citep{rivilla2017, NAZARI2024}. 

This study therefore serves multiple purposes within the framework of the ALMAGAL survey. First, it provides a detailed characterisation of the chemical inventory of one of the most chemically rich HMCs in the sample. Second, it provides a robust and reproducible template for molecular line identification and analysis that can be consistently applied to the full ALMAGAL dataset. Third, by carefully assessing line blending effects affecting GA, EG, and MF, this work ensures a reliable derivation of their physical parameters and supports forthcoming survey-wide analyses. Finally, the spatially resolved analysis of these species enables us to evaluate their ability to trace the innermost warm and dense regions of HMCs, where COM chemistry is expected to be most active.

\section{The source}

\begin{table*}[!t]
  \centering
  \caption{Properties of the core AG318$-$c9.}
  \label{tab:properties}
    \begin{tabular}{cc|cc|cc} 
        \toprule
        RA (ICRS)   & $\rm 15^h\,00'\,55\fs286$ & $F_{\rm PEAK}$ & 64.5\,$\rm mJy\,beam^{-1}$  & PA    & 42\,$\degree$ \\
        DEC (ICRS)  &  $\rm -58\degree\,58'\,52\farcs392$ & $F_{\rm INT}$ & 140.41\,mJy & $M_{\rm core}$ & 124.2\,$\rm M_{\odot}$ \\
        $D_{\rm HEL}$ & 10.4\,kpc & Deconvolved size & $\rm 0\farcs47\times0\farcs43$  & $n_{\rm H_2}$  & 17.2$\times 10^7\,\rm cm^{-3}$ \\
        $D_{\rm GAL}$ & 6.84\,kpc  & Deconvolved physical size & $4893\times4476$\,au   & $N_{\rm H2}$ & 7.35$\times 10^{24}\,\rm cm^{-2}$ \\  
        \bottomrule
    \end{tabular}
    \tablefoot{Parameters derived from \cite{almagal1} and \cite{almagal3}. $F_{\rm PEAK}$ is the background-subtracted peak flux, $F_{\rm INT}$ is the integrated flux, and PA is the position angle.}
\end{table*}

\subsection{\texorpdfstring{AG318.9477$-$00.1960}{AG318.9477-00.1960} clump}
\label{sect:source}
AG318.9477$-$0.1960 is a high-mass clump located at a heliocentric distance of $D_{\rm {HEL}}\sim\rm 10.4\pm0.4\,kpc$ and a Galactocentric distance of $D_{\rm GAL}\sim \rm 6.84\,kpc$. It is centred at $\rm \alpha_{clump}(ICRS)=15^h\,00^{m}\,55\fs293$ and $\rm \delta_{clump}(ICRS) = -58\degree\,58'\,52\farcs511$. The clump has a mass of $M_{\rm clump} \sim 5870\,M_{\odot}$, a luminosity of $L_{\rm clump}\sim 2.7\, \times 10^{5}\,L_{\odot}$, an $L/M$ $\sim 46.5\,\rm L_{\odot}/M_{\odot}$, and a systemic velocity of $V_{\rm lsr}=-34.4$\,km\,s$^{-1}$ \citep{almagal1}.

The source is associated with compact mid-IR and far-IR emission, indicating a dense and embedded environment. The radio continuum emission observed with the Australia Telescope Compact Array (ATCA) at $\rm 5.5-22.8\,GHz$ detected unresolved free-free emission with a rising spectral index of $\alpha = 0.69\pm 0.22$, which was interpreted as arising from a partially optically thick conical non-collimated radio jet \citep{purser}. The possibility that the free-free emission is associated with an HII region was dismissed based on the low radio luminosity measured at 22.8\,GHz, which is only 1.4\% of that expected for an optically thin HII region. Additional evidence of jet activity includes aligned knots of $\rm H_2$ emission \citep{Lee,debuizer2003}, a bipolar $\rm SiO$ outflow \citep{debuizer2009}, and $\rm CH_3OH$ masers at 36\,GHz and 44\,GHz \citep{voronkov}. 

\subsection{Core 9}
\label{subsect:core9}
The compact cores embedded in the ALMAGAL clumps were identified with the structure identification algorithm CUrvature Thresholding EXtractor (CuTEx; \citealt{cutex}) based on their integrated dust continuum flux. This algorithm has allowed us to identify 6348 cores in 844 of the 1013 ALMAGAL clumps \citep{almagal3}. By fitting Gaussians to each extracted core, \cite{almagal3} have estimated the dust continuum emission from the line-free channels and the size of the continuum emission. Within clump AG318.9477$-$0.1960, 19 compact cores were identified (Fig.~\ref{AG318_core9}), among which core 9 stands out as a particularly chemically rich source. We discuss the criteria used to choose this source among all the others in the ALMAGAL sample in \ref{sect:identification richest core} in detail. The properties of AG318$-$c9 are described in Table~\ref{tab:properties}.

\section{Observations} 
The ALMAGAL large project observed a total of 1013 targets using ALMA. The observations were carried out with the main 12\,m array in two different configurations: an extended configuration (C-5 and C-6), and a compact configuration (C-2 and C-3), hereafter referred to as TM1 and TM2, respectively; \citep{ALMAcorrelator}. The observations were also carried out with the 7\,m Atacama Compact Array (\cite{ACA}; hereafter 7M). We used data that combine all three configurations (7M + TM1 + TM2), yielding a synthesised beam of $\rm 0\farcs46 \times 0\farcs37$ with a PA$=7\,\degree$. For AG318$-$c9, this corresponds to physical scales of $\rm \sim 4800\,au \times 3900\,au$ along the major and minor axes, respectively.

The observations were conducted in ALMA Band~6 at $\sim\,$1.38 mm ($\sim 217$\,GHz). 
The spectral setup comprised four spectral windows (spw). We used spw0 and spw1, each covering a bandwidth of 1.87\,GHz with 3840 channels of 488\,kHz, corresponding to a spectral resolution of $\rm \approx 1.4\,km\,s^{-1}$. These spectral windows cover the frequency ranges 216.98--218.86\,GHz and 219.06--220.94\,GHz, respectively. The remaining two spectral windows (spw2 and spw3), which cover narrower bandwidths of 0.468\,GHz, were not considered here as they correspond to higher spectral resolution subwindows within the frequency ranges already covered by spw0 and spw1 and do not provide additional constraints for line identification, especially for weak species. The RMS noise of the continuum map is $\sigma \approx 0.28 \,\rm mJy\,beam^{-1}$ \citep{almagal3}. A detailed description of the ALMAGAL data-processing workflow, including calibration, data staging, continuum determination, self-calibration, imaging, and data products, is provided in \cite{almagal2}.

\section{Method}
\label{methodology}

To study the richness of the spectral line emission of each core, we extracted the spectra in the two broad spectral windows (spw0 and spw1) by integrating the emission over the ellipse that describes the size of each continuum core. The core identification procedure is described in detail in the ALMAGAL core catalogue \citep{almagal3}.

\subsection{Identification of the richest chemical core}
\label{sect:identification richest core}

We sorted the spectra of all the ALMAGAL cores using the intensity of the $K=2$ component of methyl cyanide (CH$_3$CN), where $K$ denotes the projection of the total rotational angular momentum onto the principal axis of the molecule, at 220.730\,GHz. This transition is one of the brightest in the spectral setup, is minimally affected by line blending, and is a well-established tracer of HMCs (e.g. \citealt{cesaroni_hmc, cesaroni2011}). After sorting the spectra of the ALMAGAL cores, we found that sources with weak emission in the low-$K$ components of CH$_3$CN exhibit little or no complex chemistry, whereas sources with strong low-$K$ CH$_3$CN emission display chemically rich spectra, consistent with previous studies \citep[e.g. G31.41+0.31;][]{beltran2005, guapos1}. We therefore selected sources with a very good signal-to-noise ratio (S/N) emission in the CH$_3$CN $K=2$ transition because weaker and less abundant species would otherwise not be reliably detected.

Since the forthcoming study will focus on GA, EG, and MF, we further considered the intensity of one of the brightest MF transitions in the observed band, the $\rm 17_{3,14}\rightarrow 16_{3,13}$ at 218.298\,GHz. The two selection criteria consistently place AG318$-$c9 among the cores with the highest S/N in the ALMAGAL sample, supporting its use as a template source for detailed chemical analysis.

Moreover, we selected AG318$-$c9 as a representative chemically rich core within the ALMAGAL sample based on additional criteria. First, it exhibits a high line density of approximately $\sim$80\,$\rm lines\,GHz^{-1}$ (S\'anchez-Monge et al., in prep.). This line density is comparable to that observed in the most chemically rich cores of Sgr~B2, one of the chemically richest regions in the Galaxy \citep{belloche2025}. Second, AG318$-$c9 does not show strong multiple velocity components, which greatly facilitates reliable line identification and detailed spectral analysis. Third, AG318$-$c9 is the third most massive core in the sample. The two more massive cores were excluded because the most massive core shows prominent self-absorption features, while the second core exhibits highly noisy spectra.

\subsection{Line identification}
\label{subsect: line identification}
For the line identification in AG318$-$c9, we searched for molecular species with an S/N $\geq5$ (corresponding to $\sim 1.35\,$K) to ensure reliable detections. We first fitted species expected to be bright and abundant in HMCs, such as H$_2$CO, CH$_3$CN, or CH$_3$OH, which simultaneously account for multiple of the brightest transitions in the observed frequency range. For the remaining unidentified lines, we used \texttt{Splatalogue}\footnote{\url{https://splatalogue.online/\#/advanced}} to search for additional molecular species with transitions consistent with the observed frequencies, drawing spectroscopic data from the CDMS\footnote{\url{https://cdms.astro.uni-koeln.de/classic/}} \citep{cdms2001,cdms2005, cdms2016} and the JPL\footnote{\url{https://spec.jpl.nasa.gov/ftp/pub/catalog/catdir.html}} \citep{jpl} catalogues.

To confirm or discard the identification of the different candidate species, we used the spectral analysis software Madrid data cube analysis \citep[MADCUBA \footnote{\url{https://cab.inta-csic.es/madcuba/}};][]{MADCUBA}. This software allowed us to identify and fit different species using the tool called spectral line identification and modeling (SLIM), which generates a synthetic spectrum under the assumption of local thermodynamic equilibrium (LTE) while accounting for line opacity. Given the high gas density of AG318$-$c9, $n\rm _{H_2}=1.72 \times 10^8$\,cm$^{-3}$ \citep{almagal3}, the assumption of LTE is well justified. The H$_2$ column density, $N_{\rm H_2}=7.35\times 10^{24}\,\rm cm^{-2}$ (Table~\ref{tab:properties}), was derived using $N_{\rm H_2} = \frac{4\,M_{\rm core}}{\pi\,\mu\,m_{\rm H}\,D{\rm core}^2}$, where $M_{\rm core}$ is the mass of the core, and $D_{\rm core}=6396\,\rm au$ is the geometric mean size of AG318$-$c9, as reported by \cite{almagal3}. Here, $m_{\rm H}=1.67\times 10^{-24}\,\rm g$ is the hydrogen atom mass, and $\mu=2.8$ is the mean molecular weight per particle. This $N_{\rm H_2}$ was then used to compute molecular abundances relative to H$_2$ ($X = N_{\rm species}/N_{\rm H_2}$).

The synthetic spectra were generated and controlled by input physical parameters: column density ($N$), excitation temperature ($T_{\rm ex}$), radial velocity of the source (V), full width at half maximum (FWHM), and source size ($\theta_s$). Best-fit solutions were obtained using the AUTOFIT routine, which performs a non-linear least-squares minimisation based on the Levenberg-Marquardt algorithm \citep{lm1,lm2}.

All detectable molecular species towards AG318$-$c9 were initially identified and fitted manually using MADCUBA. As we aimed to establish a framework for the chemical analysis of the full ALMAGAL sample, AG318$-$c9 was used as a testbed to develop a script that automates the line identification and fitting procedures (Appendix~\ref{macro}). This automation facilitates the evaluation of individual molecular species in the survey, improving the robustness of the derived physical parameters and their associated uncertainties, particularly for fainter species, such as EG and GA. Although we attempted to derive all the physical parameters for each species, in some cases, AUTOFIT did not converge, and certain parameters had to be fixed. In particular, $T_{\rm ex}$ was fixed to the value derived for MF ($T_{\rm ex}^{\rm CH_3OCHO}$) when only one transition was available or when the detected transitions did not span a sufficiently wide range of energies to converge in a solution. The full procedure is described in Appendix~\ref{macro}.

\begin{table*}[t]
\centering
\caption{Molecular species detected in core~9 of the AG318.9477$-$0.1960 clump.}
\begin{tabular}{c c c c c c c}
\toprule
N$^{\degree}_{\rm atoms}$ & \multicolumn{6}{c}{Simple species} \\
\midrule
2 & $\rm ^{13}CO$ & $\rm C^{18}O$ & $\rm SO$  &  $\rm ^{33}SO$ & $\rm SiO$ & $\rm SiS$ \\
3 & $\rm SO_2$ & $\rm ^{34}SO_2$ & $\rm DCN$ & $\rm O^{13}CS$ & & \\
4 & $\rm H_2CO$ & $\rm H_2^{13}CO$  & $\rm H_2C^{17}O$ & $\rm H_2CS$ & $\rm HNCO$  & $\rm H^{15}NCO$ \\
 & $\rm HN^{13}CO$ & $\rm HDCS$ & & & & \\
5 & $\rm H_2CCO$ & $\rm HC_3N$ & $\rm HC_3N, v_6=1$ & $\rm HC_3N, v_7=1$ & $\rm HC_3N, v_7=2$ & $\rm HC^{13}CCN$ \\
 & $\rm HCC^{13}CN$ & $\rm H^{13}CCCN$ & $\rm HCCC^{15}N$ & $\rm HC^{13}CCN, v_7=1$ & $\rm NH_2CN$ & $\rm H_2CNH$  \\
& $\rm t$-HCOOH & $\rm t$-H$^{13}$COOH &  &  & & \\
\midrule
 & \multicolumn{6}{c}{COMs} \\
\midrule
6  & $\rm CH_3OH$ & $\rm ^{13}CH_3OH$ & $\rm CH_3^{18}OH$ & $\rm CH_3OD$ & $\rm CH_2DOH$ & $\rm CHD_2OH$ \\
  & $\rm CH_3CN$ & $\rm CH_3^{13}CN$ & $\rm HC(O)NH_2$ & $\rm H^{13}C(O)NH_2$ & $\rm DC(O)NH_2$ & $\rm HC(O)NH_2\,v_{12}=1$ \\
7  & $\rm CH_3CHO$ & $\rm ^{13}CH_3CHO$ & $\rm c$-C$_2$H$_4$O & $\rm CH_3NCO$ & $\rm HOCH_2CN$ & $\rm C_2H_3CN$ \\
8  & $\rm CH_3OCHO$ & $\rm CH_2(OH)CHO$ & $\rm CH_3COOH$ & $\rm CH_3COOH, vt=1$ &  &  \\
9  & $\rm CH_3OCH_3$ & $\rm ^{13}CH_3OCH_3$ & $\rm C_2H_5OH$ & $\rm C_2H_5CN$ & $\rm C_2H_5^{13}CN$ & $\rm C_2H_5CN, v_{20}=1$-$A$ \\
  & CH$_3$CONH$_2$ & & & & & \\
10 & $\rm a$-(CH$_2$OH)$_2$ & $\rm g$-(CH$_2$OH)$_2$ & $\rm CH_3COCH_3$ &  &  \\
11 & $\rm C_2H_5OCHO$ &  &  &  &  &  \\
\bottomrule
\end{tabular}
\label{tab:detected_species}
\tablefoot{Species are ordered by increasing total number of atoms from left to right.}
\end{table*}

The species for which we left $T_{\rm ex}$ free generally span a wide range of energies, with minimum values typically of a few tens of kelvin and maximum values reaching several hundred kelvin. The widest $E_{\rm up}$ range was found for NH$_2$CN (579\,K), and the smallest range was found for CH$_3$COOH (57\,K).

Since EG, GA, and MF have clearly been detected in AG318$-$c9, we also generated $N$ and $T_{\rm ex}$ maps (Sect.~\ref{subsect:physical_maps}). Pixel-by-pixel fitting of selected transitions was performed (Appendix~\ref{sec:maps}), allowing us to better constrain the physical conditions and chemical structure of the core.

\section{Spectral line analysis}
We present the complete chemical inventory analysed in AG318$-$c9 derived from the ALMAGAL frequency setup. To place these results, we compare them with those of the well-studied HMC G31.41+0.31.
\subsection{Molecular inventory}
\label{sect:results}
\begin{figure*}[!h]
    \resizebox{\hsize}{!}{\includegraphics{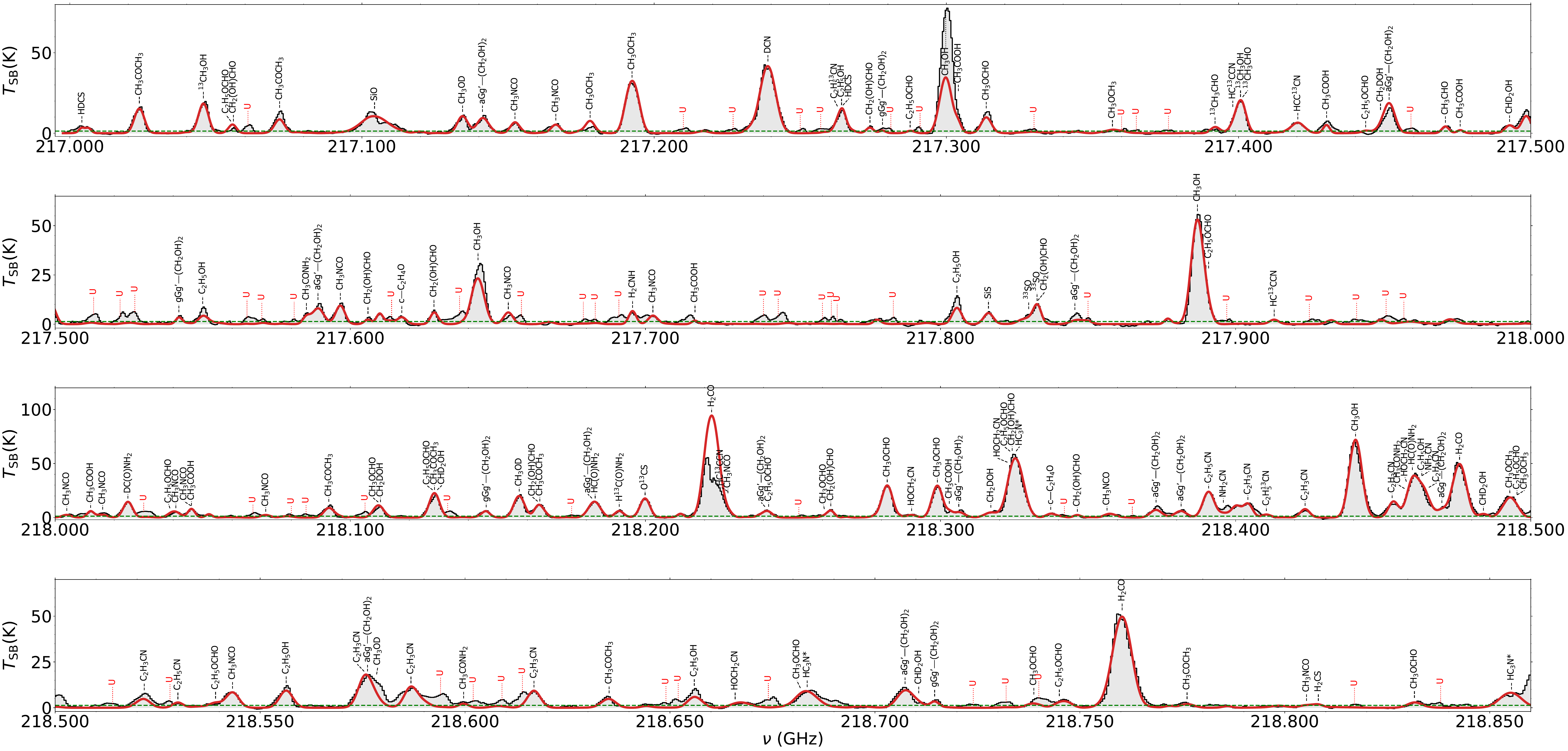}}
    \centering
    \caption{Spectra of spw0 towards core 9 of the AG318.9477$-$00.1960 clump. The black histogram and its grey shadow are the observational spectrum, and the red curve indicates the LTE fit after taking into account the contribution of all the species (see Table~\ref{Table:physical_parameters} and its complete version in Appendix~\ref{Tableapp:physical_parameters}). The horizontal dashed green line indicates the $\rm 5\sigma \ \approx 1.35\,K$ level, with $\rm \sigma$ being the noise ($\rm \sigma \sim 0.28\,K$). Only transitions contributing at $\geq5\sigma$ are labelled. The vertical dashed black lines indicate the position where $\rm \nu$ was converged. U labels indicate unidentified transitions.}
    \label{fig:madcuba_spw0}
\end{figure*}

\begin{figure*}[!h]
    \resizebox{\hsize}{!}{\includegraphics{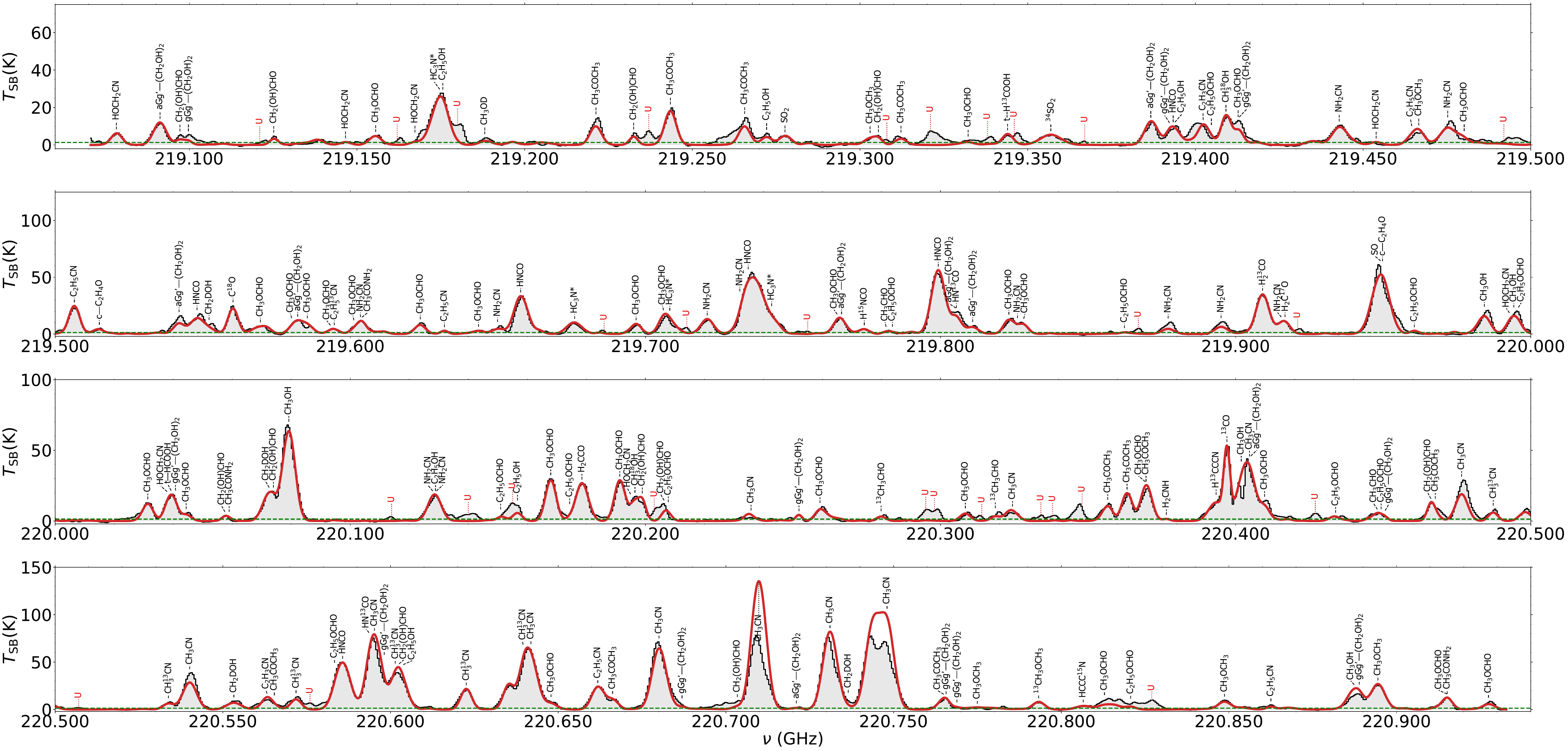}}
    \centering
    \caption{Spectra of spw1 towards core 9 of the AG318.9477$-$00.1960 clump. The black histogram and its grey shadow are the observational spectrum, and the red curve indicates the LTE fit after taking into account the contribution of all the species (see Table~\ref{Table:physical_parameters} and its complete version in Appendix~\ref{Tableapp:physical_parameters}). The horizontal dashed green line indicates the $\rm 5\sigma \ \approx 1.35\,K$ level, with $\rm \sigma$ being the noise ($\rm \sigma \sim 0.28 K$). Only transitions contributing at $\geq5\sigma$ are labelled. The vertical dashed black lines indicate the position where $\rm \nu$ was converged. U labels indicate unidentified transitions.}
    \label{fig:madcuba_spw1}
\end{figure*}

The analysis of the spectra reveals molecular emission from a total of 65 different species (Table~\ref{tab:detected_species}). This underlines the chemically rich environment in AG318$-$c9. Of these species, 31 correspond to rotational transitions of main species, and the remaining 34 species arise from isotopologues and vibrationally excited species. Figures~\ref{fig:madcuba_spw0} and \ref{fig:madcuba_spw1} show the observed spw0 and spw1 spectra together with the global LTE fit, which accounts for the contribution of all identified species. Zoomed-in versions of the spectra, presented in Appendix~\ref{zoom spw}, illustrate the quality of the fits for weaker transitions. Despite the large number of detected species, not all observed spectral features could be reliable identified. A total of 91 lines remained unassigned and are labelled unidentified (U) in the spectra. Although the most recent CDMS and JPL catalogues were used, some of these unidentified features may correspond to transitions that are not yet included in current laboratory spectroscopic databases. 

Overall, the identified molecular inventory reveals a chemically rich environment dominated by oxygen-bearing species (44), followed by nitrogen-bearing (28), sulphur-bearing (8), and silicon-bearing (2), with COMs accounting for more than half of the detections (33). In particular, GA, EG, and MF were identified through multiple transitions with good S/N ratios ($\geq 5$), of which 3, 8, and 14 transitions, respectively, were unblended. Additional heavy COMs, such as acetone ($\rm CH_3COCH_3$) and ethyl formate ($\rm C_2H_5OCHO$), are also present. 

Remarkably, this rich chemical inventory is revealed by covering a narrow spectral setup of $\rm 4\,GHz$ and with short integration times. The number of different species detected in AG318$-$c9 is comparable with those detected towards typical chemically rich HMCs such as G31.41+0.31, Sgr B2, and IRAS 16293$-$2422. However, the inventory of species in these well-known  HMCs has been built through large unbiased spectral surveys: e.g., the GUAPOS survey (G31.41+0.31 Unbiased ALMA sPectral Observational Survey; \citealt{guapos1}) and the EMoCA survey (Exploring molecular complexity in the Galactic Centre with ALMA; \citealt{Belloche_2017}), which both span nearly the entire ALMA Band 3 ($\sim 32$\,GHz), and the PILS survey (The ALMA Protostellar Interferometric Line Survey; \citealt{pils}) from 329 to 363\,GHz ($\rm \sim 34\,GHz$), allowing a more complete census of molecular species. All this suggests that AG318$-$c9 is a chemically reference HMC within the ALMAGAL survey that might be as rich as other well-known and studied sources.

The detected species span a wide range of molecular complexity, from simple diatomic molecules such as the isotope of carbon monoxide $\rm ^{13}CO$ to molecules with 11 atoms such as ethyl formate ($\rm C_2H_5OCHO$).
The complete list of identified species and their correspondent transitions is available in a repository table\footnote{The complete table of molecular transitions used for the MADCUBA fits is available at \url{https://github.com/JofreAllande/ALMAGAL/}}. 
Together, this inventory supports the classification of AG318$-$c9 as a chemically rich HMC.

\renewcommand*{\arraystretch}{1.2}
\begin{table*}[!b] 
\centering
\caption{Results of the SLIM fits of the species analysed towards AG318$-$c9. Only a portion of the table is shown here. The complete table is available in Appendix~\ref{Tableapp:physical_parameters}}
\begin{tabular}{@{\extracolsep{\fill}}ccccccc@{\extracolsep{\fill}}}
 \hline
{Species} & {$T_{\rm ex}$ (K)} & {$N$ ($\times 10^{16}$ cm$^{-2}$)} & {$X$ ($\times 10^{-9}$)} & {FWHM (km s$^{-1}$)} & {$V$ (km s$^{-1}$)} & {N$\degree$ transitions} \tabularnewline 
\hline 
$^{13}$CO$^{(a)}$ & $167$ & $42 \pm 3$ & $57 \pm 4$  & $3.2 \pm 0.3$ & $-31.59 \pm 0.12$ & 1\tabularnewline 
C$^{18}$O & $167$ & $30.4 \pm 1.3$ & $41.3 \pm 1.7$ & $5.4 \pm 0.3$ & $-33.66 \pm 0.11$ & 1\tabularnewline 
SO$^{(b)}$ & & $12.2 \pm 1.1$ & $16.5 \pm 1.4$ & $9.0 \pm 0.3$ & $-33.34 \pm 0.16$ & 1\tabularnewline 
$^{33}$SO & $167$ & $0.136 \pm 0.010$ & $0.185 \pm 0.014$ & $5.0$ & $-33.9 \pm 0.3$ & 4 \tabularnewline
SiO & $167$ & $0.054 \pm 0.002$ & $0.073 \pm 0.003$ & $15.0 \pm 0.6$ & $-32.9 \pm 0.3$ & 1 \tabularnewline 
SiS & $167$ & $0.039 \pm 0.003$ & $0.053 \pm 0.004$ & $5.2 \pm 0.5$ & $-31.9 \pm 0.2$ & 4\tabularnewline 
SO$_2$$^{(b)}$ &  & $24.3 \pm 1.4$ & $33.1 \pm 1.9$ & $5.7 \pm 0.6$ & $-36.0 \pm 0.2$ & 2\tabularnewline 
$^{34}$SO$_2$ & $167$ & $0.77 \pm 0.07$ & $1.05 \pm 0.10$ & $10.0$ & $-36.4 \pm 0.5$ & 2\tabularnewline 
DCN & $167$ & $0.130 \pm 0.002$ & $0.178 \pm 0.003$ & $8.13 \pm 0.15$ & $-34.35 \pm 0.07$ & 1 \tabularnewline 
$\vdots$ & $\vdots$ & $\vdots$ & $\vdots$ & $\vdots$ & $\vdots$ & $\vdots$ \tabularnewline 
\hline 
\end{tabular}
\tablefoot{
 The derived physical parameters ($T_{\rm ex}$, $N$, $X=N_{\rm species}/N_{\rm H_2}$, FWHM, and $V$) are listed along with their associated uncertainties. The last column indicates the number of transitions used to fit the spectrum. The parameters we fixed during the fitting procedure are reported without uncertainties.
When $T_{\rm ex}$ is fixed, $T_{\rm ex}^{\rm CH_3OCHO}=167\,\rm K$ is assumed (see Sect.\,\ref{macro}). HC$_3$N* represents the group composed of
HC$_3$N, HC$_3$N ($v_6=1$), HC$_3$N ($v_7=1$), and HC$_3$N ($v_7=2$).
\tablefoottext{a}{Species exhibiting a unique self-absorbed transition,
resulting in a lower limit on the column density.}
\tablefoottext{b}{The column density was estimated from the less abundant isotopologues using the isotopic ratios $\rm ^{12}C/^{13}C = 46.6\pm10$ \citep{Yan_2019}, $\rm ^{32}S/^{34}S = 20.7\pm1.2$ \citep{yan2023}, and $\rm ^{32}S/^{33}S = 88.9\pm7.7$ \citep{yan2023}.}}
\label{Table:physical_parameters}
\end{table*}
\renewcommand*{\arraystretch}{1.0} 

For all identified species, we derived the column density, excitation temperature, velocity, line width, and abundance (Table~\ref{Table:physical_parameters}), following the procedure described in Appendix~\ref{macro}. The excitation temperatures of the detected species are in the range 51--440\,K, while the column densities are 10$^{13}$--$10^{18}$\,cm$^{-2}$, and their relative abundances range from 10$^{-11}$ to $10^{-7}$. For species for which isotopologues were also detected ($\rm H_2CO$, $\rm SO_2$, $\rm SO$, $\rm HC_3N$, $\rm HNCO$, $\rm HC(O)NH_2$, $\rm CH_3OH$, $\rm CH_3CN$, $\rm t-HCOOH$, $\rm CH_3OCH_3$, $\rm CH_3CHO$, and $\rm C_2H_5CN$), the column densities of the main isotopologues were obtained by scaling those of the isotopologues by the measured isotopic ratios. These ratios were calculated taking the Galactocentric distance ($D_{\rm GAL}$) of AG318$-$c9 into account, which is 6.84\,kpc (calculated from the $D_{\rm HEL}$ of 10.4 kpc; \citealt{almagal1}). Specifically, we adopted $\rm ^{ 12}C/^{13}C = 46.6\pm10$ \citep{Yan_2019}, $\rm ^{32}S/^{33}S = 88.9\pm7.7 $, and $\rm ^{32}S/^{34}S = 20.7\pm1.2 $ \citep{yan2023}.

Overall, the global fit reproduces the observed spectra well in most frequency ranges, with the exception of some transitions of $\rm CH_3CN$, $\rm H_2CO$, and $\rm ^{13}CO$, which are not fully reproduced by the synthetic spectrum and present overestimated emission. This behaviour arises because the $\rm CH_3CN$ transitions with $K=0,1,2$ are optically thick and show absorption features, as is also the case for the $\rm H_2CO$ $\rm 3_{0,3}\rightarrow 2_{0,2}$ transition at 218.222\,GHz. The $\rm CH_3CN$,$K=3$ ladder belongs to the A symmetry species associated with the internal rotation of the $\rm CH_3$ group, whereas most other $K$ ladders are of E symmetry. Since A and E species do not interconvert, differences in their populations or traced physical conditions prevent a single LTE model from reproducing all $K$ ladders simultaneously, leading to an overestimation of the $K = 3$. In addition, the $\rm ^{13}CO$ emission is spatially extended and is therefore filtered out by the interferometer, resulting in missing flux.

The detection of multiple COMs, and in particular, of EG and GA, constitutes a key result of this study, placing this source among the list of cores exhibiting signatures of COMs with potential prebiotic relevance. GA has been reported in only a limited number of sources to date (e.g. \citealt{Hollis2000, Beltran_2009, Jorgensen_2012, coccoa}), and its secure detection towards AG318$-$c9 reinforces the view that this core hosts physical and chemical conditions that are favourable for prebiotic chemistry, including grain-surface processes and warm-up timescales enabling ice sublimation.

\subsection{Comparison with the hot core G31.41+0.31}
\label{sect:g31vsg318}
To place our results in context, we compared the chemical reservoir of AG318$-$c9 with that of the well-studied HMC G31.41+0.31 (G31 hereafter). G31 is the first core in which GA was detected outside the Galactic centre \citep{Beltran_2009}, and it is one of the first cores in which EG was found \citep{rivilla2017}. This HMC was also the target of the GUAPOS project, a spectral survey of the entire ALMA band 3 ($84.05-115.91\rm\,GHz$; \citealt{guapos1}). This HMC is located at a $D_{\rm HEL}$ of $\rm 3.75\,kpc$ \citep{Immer} with a $L=\rm 4.4 \times 10^4\,L_{\odot}$ \citep{osorio} and a gas mass $M\sim70\,M_{\odot}$\,\citep{massG31}. By contrast, AG318$-$c9 is significantly more distant, at a heliocentric distance of $\rm \sim 10.4\,kpc$, and its estimated gas mass is $M\sim120\,M_{\odot}$ \citep{almagal3}. However, this mass is likely an overestimate because it was obtained by assuming a uniform core temperature of 49\,K, whereas our $T_{\rm ex}$ maps reveal a temperature gradient in the core with temperatures significantly $>$49\,K (Sect.~\ref{subsect:physical_maps}), implying a lower total gas mass. Since a core-specific luminosity is unavailable, we estimated it based on the host clump luminosity ($\rm 2.7\times10^5\,L_{\odot}$), which is fragmented into 19 cores \citep{almagal3}. We apportioned the clump luminosity to AG318$-$c9 by scaling it by the ratio of its peak flux ($F_{\rm peak_{core}}$) to the sum of all the core peak fluxes ($F_{\rm peak_{all\ cores}}$), leading to $L_{\rm core}\approx 1.2 \times 10^5\,L_{\odot}$, which is almost three times higher than the luminosity of G31. Although we would like to interpret the evolutionary stages of the two HMCs based on their L/M ratios, this approach is subject to significant uncertainties. We note that the interpretation of the evolutionary stage based on global indicators such as the luminosity-to-mass ratio is not straightforward in AG318$-$c9. The core mass is likely overestimated, while the luminosity is indirectly inferred from the host clump, introducing additional uncertainties. On the other hand, the mass of G31 should be considered as a lower limit \citep{cesaroni_2019}. Additionally, in high-mass star-forming regions, L/M may not provide an unambiguous evolutionary indicator due to the effects of fragmentation, temperature gradients, and internal source structure \citep{motte2018}.

\begin{figure}[h!]
    \centering
    \includegraphics[width=\linewidth]{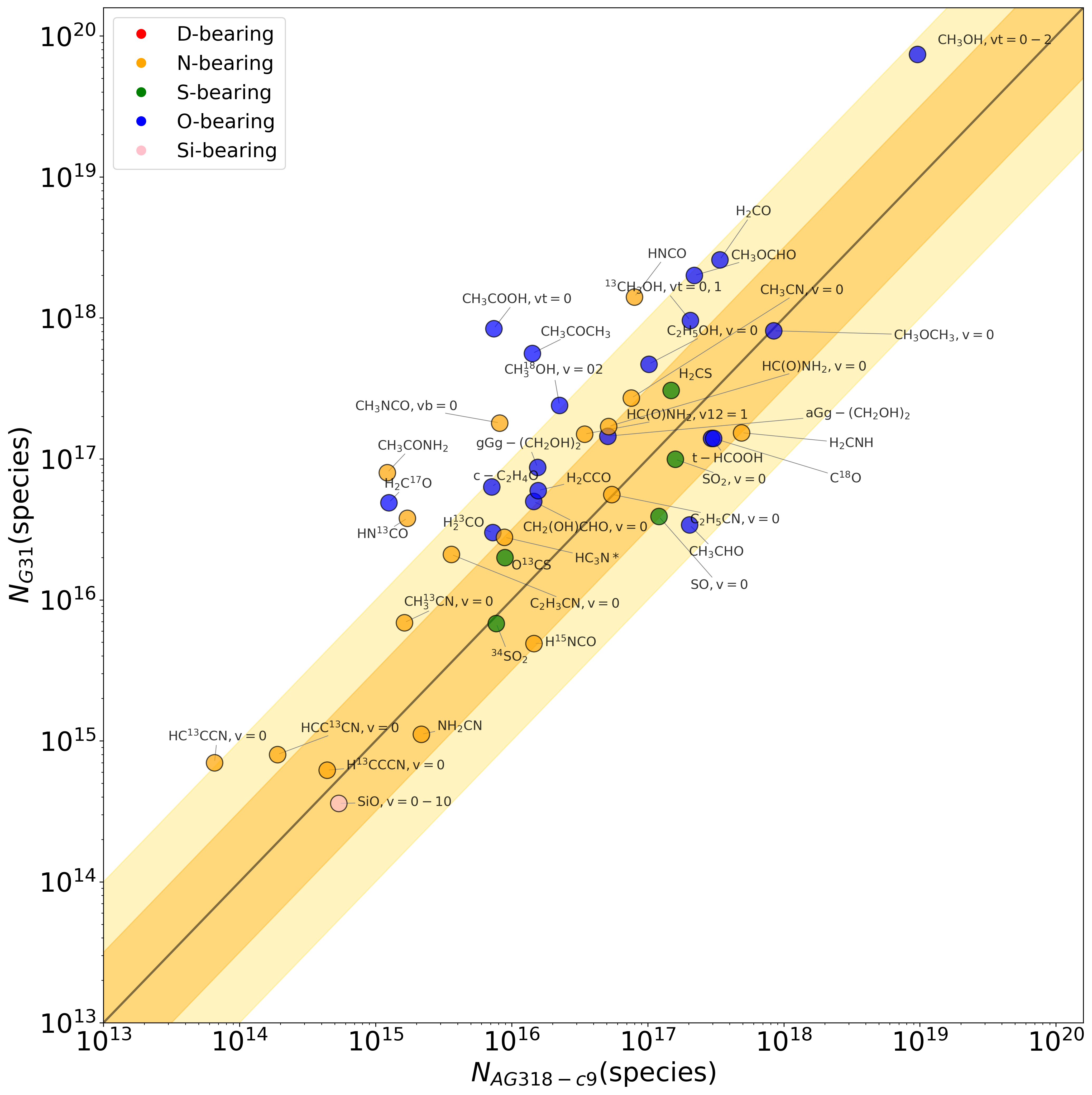}
    \caption{Column density comparison for all common species between AG318$-$c9 (x-axis) and G31 (y-axis). The solid black line indicates the 1:1 relation. The dark and light orange shaded regions represent deviations within 0.5\,dex and 1.0\,dex from equality, respectively.}
    \label{fig:plot_species_G31vsG318}
\end{figure}

The GUAPOS survey \citep{guapos1, guapos2,guapos3, guapos4, guapos5, GUAPOS6} explored the chemical richness of G31, detecting numerous O- and N-bearing COMs as formamide (HC(O)NH$_2$), methyl isocyanate (CH$_3$NCO), and ethyl isocyanide (C$_2$H$_5$NC). A one-to-one comparison of the column density for species detected in the two sources is presented in Fig.~\ref{fig:plot_species_G31vsG318}, where the spectra are extracted over comparable spatial scales ($\sim 6400$\,au for AG318$-$c9 and $\sim4500$\,au for G31), indicating that the sampled regions are broadly similar. A clear trend emerges: although G31 has a lower luminosity and mass, the column densities of the species detected towards this HMC are systematically higher than those estimated towards AG318$-$c9 for most of the species. This discrepancy is particularly pronounced for O-bearing COMs, such as $\rm CH_3COOH$, $\rm CH_3COCH_3$, and $\rm H_2C^{17}O$, which show significantly higher $N$ values in G31. Only two species present enhanced $N$ in AG318$-$c9, which are $\rm t-HCOOH$, and $\rm C^{18}O$. N-bearing species also show higher values in G31, but not as high as O-bearing species. S- and Si-bearing species show column densities that are more comparable, clustering closer to the 1:1 correlation line (i.e. within $\sim$ 1 dex). We found 11 exceptions to the general trend, in which species present enhanced $N$ in AG318$-$c9: CH$_3$CHO, t-HCOOH, C$^{18}$O, H$_2$CNH, H$^{15}$NCO, NH$_2$CN, SO$_2$, SO, $^{34}$SO$_2$, SiO, and CH$_3$OCH$_3$.

The observed chemical segregation, characterised by a pronounced excess of O-bearing COMs in G31, strongly suggests a more efficient thermal sublimation of the ice mantles in G31 \citep{guapos3}. This implies that G31 is either intrinsically warmer or at a more advanced evolutionary stage than AG318$-$c9, having had more time to release and accumulate a larger fraction of these species into the gas phase.

N-bearing species have more diverse formation pathways, including active gas-phase chemistry, and require a longer timescale to be synthesised in the cold cloud \citep{Lee2020}. Their abundances are also sensitive to environmental factors such as the UV radiation field, shocks, cosmic-ray ionisation rate, and local N/O elemental ratios. This may explain why their column densities are more comparable between the two sources. Among the exceptions, the higher $N$ of the prebiotic species cyanamide (NH$_2$CN) or the methanimine (H$_2$CNH) in AG318$-$c9 is particularly noteworthy, suggesting that the physical conditions or precursor availability in AG318$-$c9 are uniquely favourable for their production. Since D-bearing species are thought to be formed more efficiently, at the low temperatures ($<20$\,K) of the preceding cold core phase, their detection in the hot gas of AG318$-$c9 implies that they were formed on the ice grains and subsequently sublimated \citep{caselli_ceccarelli}. However, G31 also presents some of the D-bearing species found in AG318$-$c9 (V. Rivilla, priv. comm.). This suggests that the two cores share a similar chemical heritage from their cold pre-stellar phase.

The emission of Si- and S-bearing species is usually associated with shocks, as these species would be released from the dust mantles by sputtering. The fact that the column density of these species in both HMCs is comparable suggests that both cores are associated with shocks, likely produced by molecular outflows. For G31, \cite{beltran2022b} has detected at least six outflows in the core, while for the AG318.9477$-$00.1960 clump, a preliminary analysis suggests that this core is associated with at least two outflows (M. Benedettini, priv. comm.).

\begin{figure*}[h!]
    \centering
    \includegraphics[width=\linewidth]{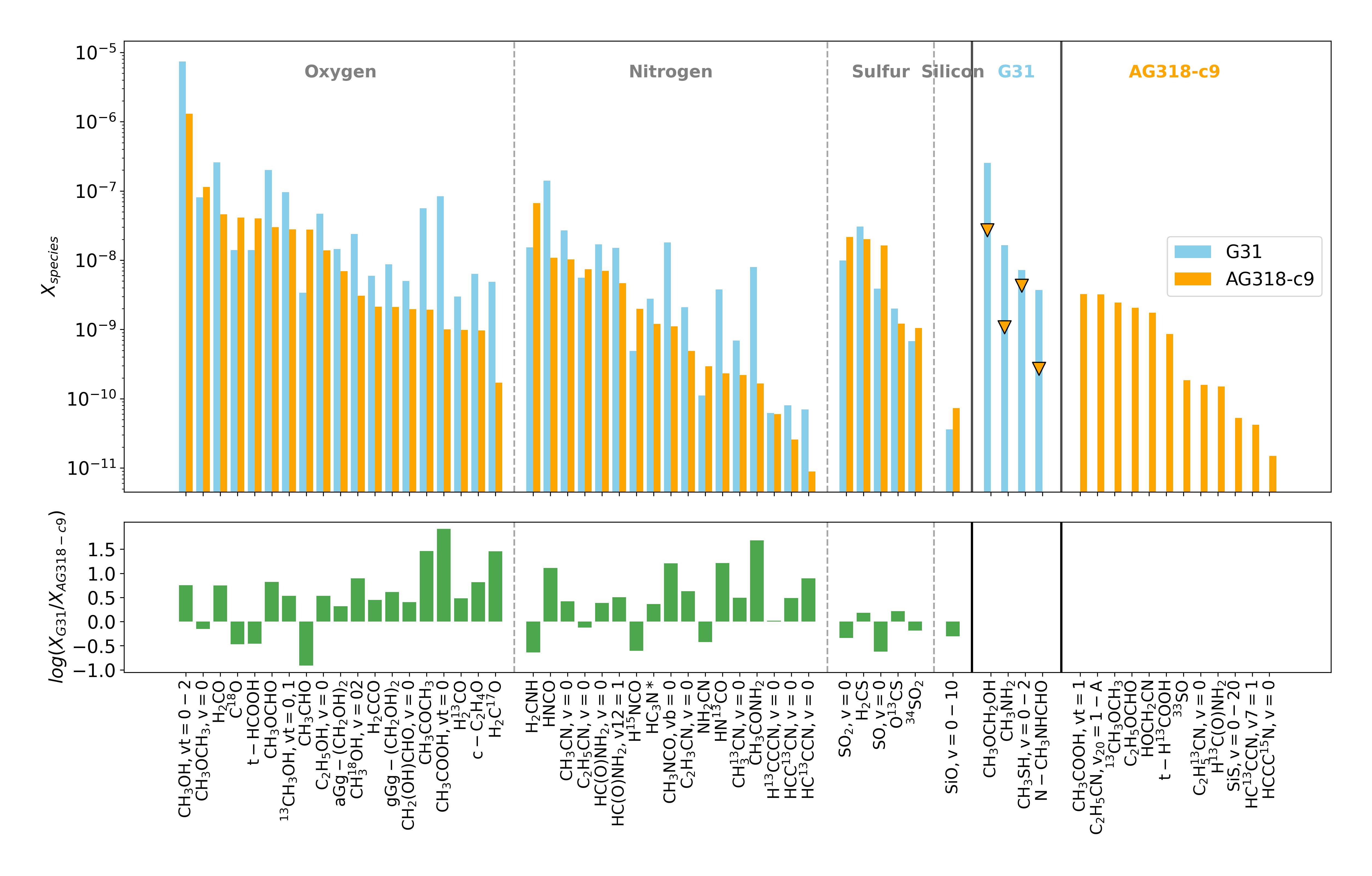}
    \caption{Top: Comparison of molecular abundances ($X = N/\rm H_2$) towards G31 (blue bars) and AG318$-$c9 (orange bars), sorted by molecular groups and by the abundance of AG318$-$c9 inside each group. The species are grouped on the x-axis according to their detection status. Left: Detected in both sources. Middle: Detected in G31 alone. The orange triangles represent the corresponding upper limits for AG318$-$c9. Right: Reported in AG318$-$c9. Bottom: Logarithm of the ratio of the abundances of the two sources. HC$_3$N* represents the group composed of HC$_3$N, HC$_3$N ($v_6=1$), HC$_3$N ($v_7=1$), and  HC$_3$N ($v_7=2$).}
    \label{fig:histogram_species_G31vsG318}
\end{figure*}

In Fig.~\ref{fig:histogram_species_G31vsG318} we show a comparative histogram of the molecular abundances ($X=N_{\rm species}/N_{\rm H_2}$) for the two sources. We also included the species detected exclusively in G31 and those that are detected in AG318$-$c9, but were not reported for G31 to date. To ensure a fair comparison, we restricted this selection to species covered by the spectral setups of both surveys. For this reason, we excluded the following molecular species: NS, OCS, NH$_3$, NH$_2$D, N$_2$H+, HCN, CO, CS, CH$_3$CCH, CH$_3$NC, s-C$_2$H$_5$CHO, and $^{13}$CO. In addition, a direct comparison with the GUAPOS survey is not yet possible because their full deuterated inventory has not yet been published, and for this reason, DCN, HDCS, CH$_3$OD, CH$_2$DOH, CHD$_2$OH, and DC(O)NH$_2$ are not included in Figs.~\ref{fig:plot_species_G31vsG318} and \ref{fig:histogram_species_G31vsG318}.

In addition to the enhanced column densities reported above, the comparison based on molecular abundances (Fig.~\ref{fig:histogram_species_G31vsG318}) indicates that C$_2$H$_5$CN is comparatively more abundant in AG318$-$c9. Moreover, a distinct chemical divergence can be noted: G31 features complex N-bearing COMs such as acetamide (CH$_3$CONH$_2$) and methylamine (CH$_3$NH$_2$). The higher $N$ of acetamide is particularly significant because it contains the peptide-like linkage (-NH-C=O), a key component for prebiotic chemistry \citep{guapos2}. The molecular inventory of AG318$-$c9 is distinguished by the presence of O-bearing species that have not been reported in G31 so far, most notably, ethyl formate (C$_2$H$_5$OOCH), along with others such as HOCH$_2$CN. The detection of ethyl formate in AG318$-$c9, together with its apparent absence in the G31 survey suggests a chemistry dominated by the thermal sublimation of O-rich ice mantles, releasing these O-bearing COMs into the gas phase \citep{Belloche_2009}.

This comparison suggests that the two sources might correspond to different evolutionary stages. AG318$-$c9 might represent a younger HMC, where the chemistry is still dominated by the recent sublimation of ice mantles rich in ester-bearing species. In contrast, G31 appears to be a more evolved environment in which this initial chemical reservoir has likely been processed by warm gas-phase reactions. 

Although the overall lower abundances in AG318$-$c9 are consistent with a less chemically processed source, abundance differences alone do not uniquely constrain the relative evolutionary stage of the two cores. Variations in temperature structure, radiation field, source geometry, or chemical pathways might also contribute to the observed differences \citep{ossenkopf2013, heays2017}. Therefore, the scenario in which AG318$-$c9 is younger than G31 should be regarded as plausible rather than definitive.
These differences indicate distinct chemical evolutionary pathways, although detailed chemical modelling is required to draw firm conclusions.

\section{Glycolaldehyde, ethylene glycol, and methyl formate}

We focused on the identification of GA, EG, and MF emission towards AG318$-$c9, paying particular attention to line-blending effects that might affect their detection.  GA stands out as a key prebiotic species, being the simplest sugar-like monosaccharide molecule that plays a fundamental role in sugar-formation pathways, leading to the formation of essential biomolecules such as glyceraldehyde and ribose, the backbone of nucleic acids. GA has been detected in different star-forming regions in the ISM, including the first detection in Sgr B2(N) in the Galactic centre \citep{Hollis2000}, the HMC G31.41+0.31 \citep{Beltran_2009}, and the solar-type protostar IRAS 16293-2422 \citep{Jorgensen_2012}. 

The heavy COM EG, with ten atoms, and the fully reduced form of GA where the aldehyde group has been reduced to an alcohol, although not directly prebiotic, is chemically linked to GA, and they may both share the same formation pathways in the ISM \citep{Sorrell_2001,fedoseev2015,coutens2018,garrod2022}. EG was first found in the ISM towards the Galactic centre \citep{hollis2002_EG} in HMCs such as G31.41 + 0.31 \citep{rivilla2017} or G34.3 + 0.1 \citep{brouillet_EG} and in hot corinos such as NGC 1333-IRAS 2A \citep{Maury_EG}. 

Finally, MF, although not chemically linked through the same formation pathways, can offer complementary constraints on the chemistry of GA and EG. All three species are expected to trace similar physical conditions \citep{Simons2020}.

\subsection{Glycolaldehyde}
\label{subsection:GA}

\begin{figure}[h]
    \centering
    \includegraphics[width=0.99\linewidth]{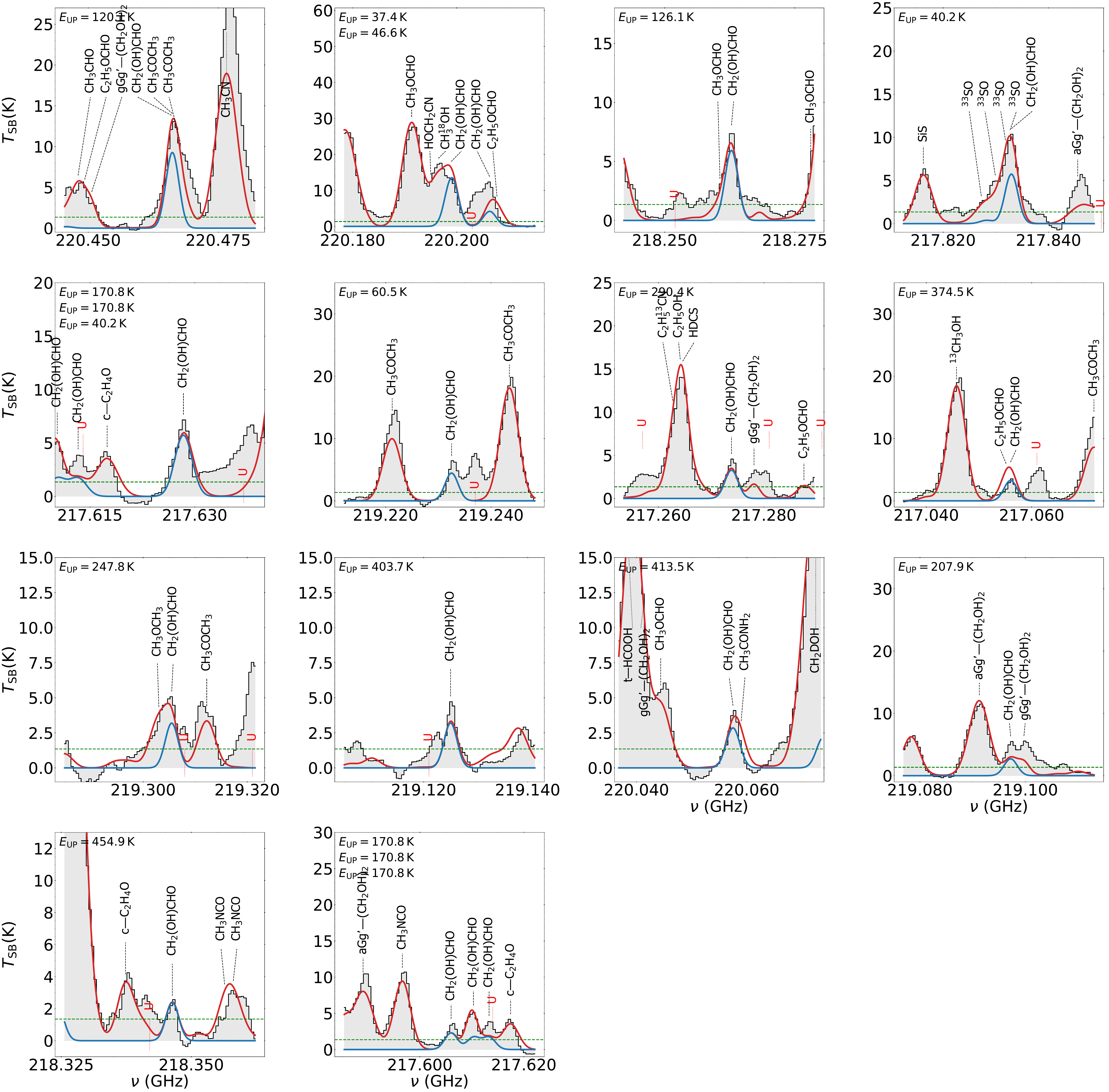}
    \caption{Selection of GA transitions observed and fit (see Sect.~\ref{subsection:GA}). The black histogram and its grey shadow are the observational spectrum, and the red curve is the cumulative LTE fit considering all detected species. The blue curve is the fit of the individual transitions. The horizontal dashed green line indicates the $\rm 5\sigma$ level, corresponding to $\rm \sim1.35\,K$. Only transitions contributing at $\geq5\sigma$ are labelled. The plots are sorted by decreasing intensity of the GA transitions (blue line). The upper left corner of each panel indicates the $E_{\rm UP}$ of the corresponding GA transition. When multiple GA transitions are shown in the same panel, the $E_{\rm UP}$ values are listed from top to bottom following the left-to-right order of the transitions.}
    \label{fig:ga_transitions}
\end{figure}

In Fig.~\ref{fig:ga_transitions} we show all 19 transitions of GA, clearly above S/N $\geq 5$. The transition $\rm 20_{2,18}\rightarrow 19_{3,17}$ is the brightest and has the highest opacity ($\tau = 0.061$), confirming that all the GA transitions are optically thin. Unlike for EG and MF (Sects.~\ref{subsection:EG} and \ref{subsection:MF}), in the available ALMAGAL frequency coverage, there are few unblended transitions of GA. There are three unblended transitions: $\rm 9_{5,5}\rightarrow 8_{4,4}$ at 217626.13\,MHz, $\rm 31_{4,27}\rightarrow 31_{3,28}$ at 217271.61\,MHz, and $\rm 20_{3,17}\rightarrow 19_{4,16}$ at 218260.59\,MHz. Most of the blending of GA is with $\rm CH_3COCH_3$, $\rm CH_3^{18}OH$, $\rm CH_3OCH_3$, and $\rm CH_3NCO$ transitions, whose contribution is included in the global red fit in Fig.\,\ref{fig:ga_transitions}, and it explains the remaining features of the observed spectrum. The fitted transitions span a broad range of energies ($E_{\rm up}$ = 26--316\,cm$^{-1}$ or 37--450\,K).

Although SLIM converged to a solution for GA without the need to fix any parameter, we fixed the FWHM to $4.0$\,km\,s$^{-1}$, as the line width obtained from the unconstrained fit might be visually refined to match the observed spectra better. The resulting fit of the detected transitions of GA indicates $T_{\rm ex} = 168\pm 13\,$K, $N=(1.45\pm 0.12)\times 10^{16}\,\rm cm^{-2}$ and $V=-36.89 \pm 0.1\,\rm km\,s^{-1}$.

\subsection{Ethylene glycol}
\label{subsection:EG}
Ethylene glycol is a heavy COM and the fully reduced alcohol form of GA. The presence of both species, which are thought to be chemically linked through common formation pathways \citep{coutens2018, Simons2020, garrod2022}, confirms the complexity of the chemistry in  AG318$-$c9. EG can adopt different conformations because torsions around its C-C bond and its two C-O bonds lead to four forms with an anti-arrangement (a-EG) of the hydroxyl groups (-OH), and six with a gauche-arrangement (g-EG) \citep{conformational_EG,rivilla2017}. We focused on the a-EG conformer because it shows a significantly larger number of transitions within the ALMAGAL spectral setup compared to the g-EG conformer. The latter shows many blended lines, and the few unblended transitions have similar energies, $E_{\rm up}\approx 130\,\rm K$, so that it is difficult to properly constrain the physical parameters. 

\begin{figure}[h!]
    \centering
    \includegraphics[width=0.99\linewidth]{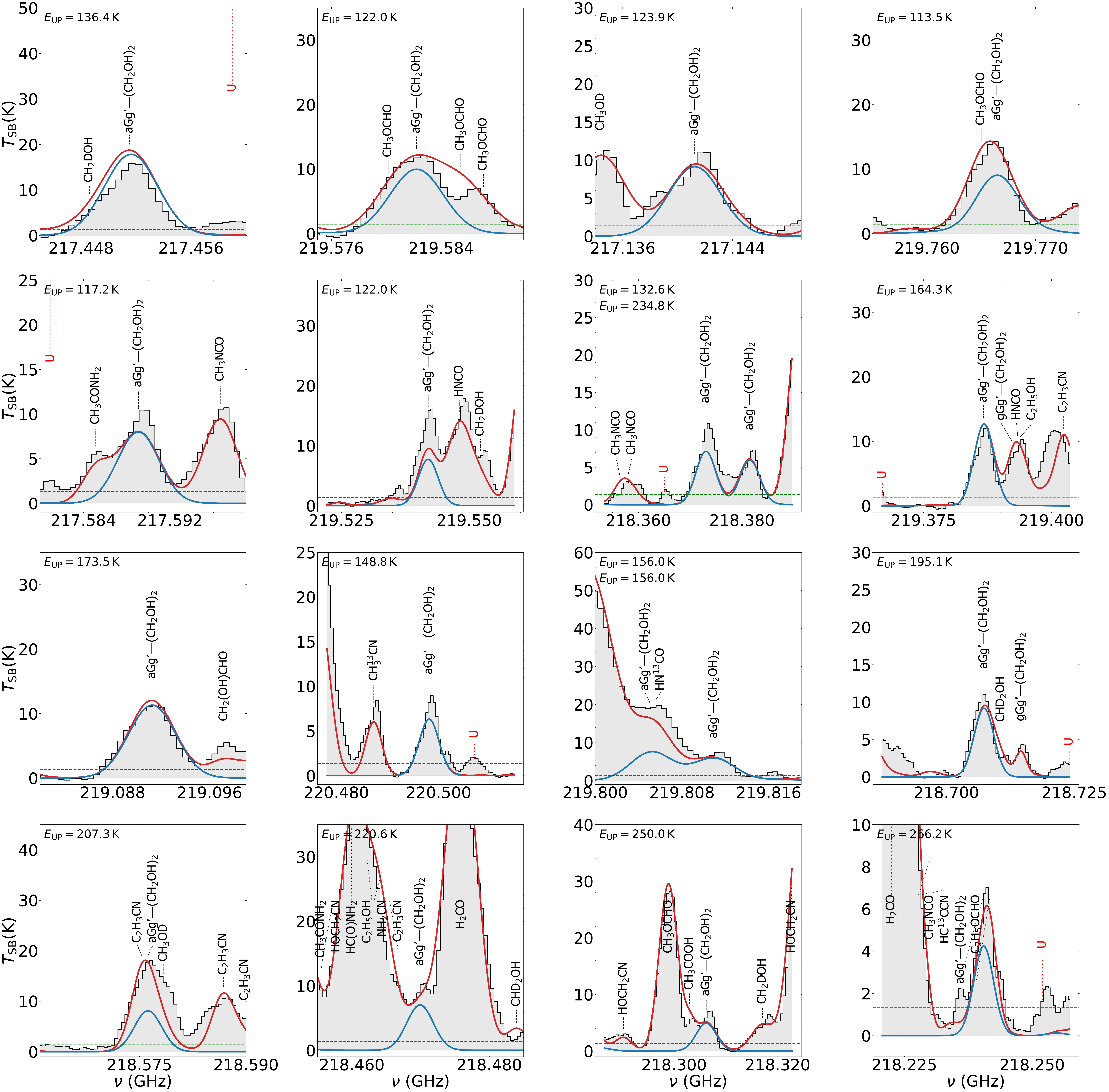}
    \caption{Same as Fig. \ref{fig:ga_transitions}, but for a-EG, as described in Sect. \ref{subsection:EG}.}
    \label{fig:eg_transitions}
\end{figure}

Figure~\ref{fig:eg_transitions} shows selected lines corresponding to a-EG, eight of which are unblended and strong. Other transitions are more blended, mainly with MF, $\rm CH_3NCO$, $\rm HNCO$, $\rm HN^{13}CO$, $\rm C_2H_5OH$, and $\rm CHD_2OH$. The $\rm 24_{1,24}\rightarrow 23_{1,23}$ transition at 217450\,MHz is the brightest and the most optically thick with $\tau = 0.046$. To fit the EG emission, we took 37 different transitions into account, which span a range of energies of E$_{\rm up}$ = 79--389\,cm$^{-1}$ or 114--580\,K. Fig.~\ref{fig:eg_transitions}shows that the LTE fit, with the joint contribution of all blending species, is able to properly explain the observed emission.

For a-EG, SLIM converged to a solution for a-EG without the need to fix any parameters. The resulting fit of the detected transitions of a-EG indicates $T_{\rm ex}$ = $240\pm 30$\,K,  $N=(5.1\pm 0.6)\times 10^{16}$\,cm$^{-2}$, $\rm FWHM = 6.19\pm 0.19 \, km\, s^{-1}$, and $V=-35.9 \pm 0.08\,\rm km\,s^{-1}$.

\subsection{Methyl formate}
\label{subsection:MF}
\begin{figure}[h]
    \centering
    \includegraphics[width=0.99\linewidth]{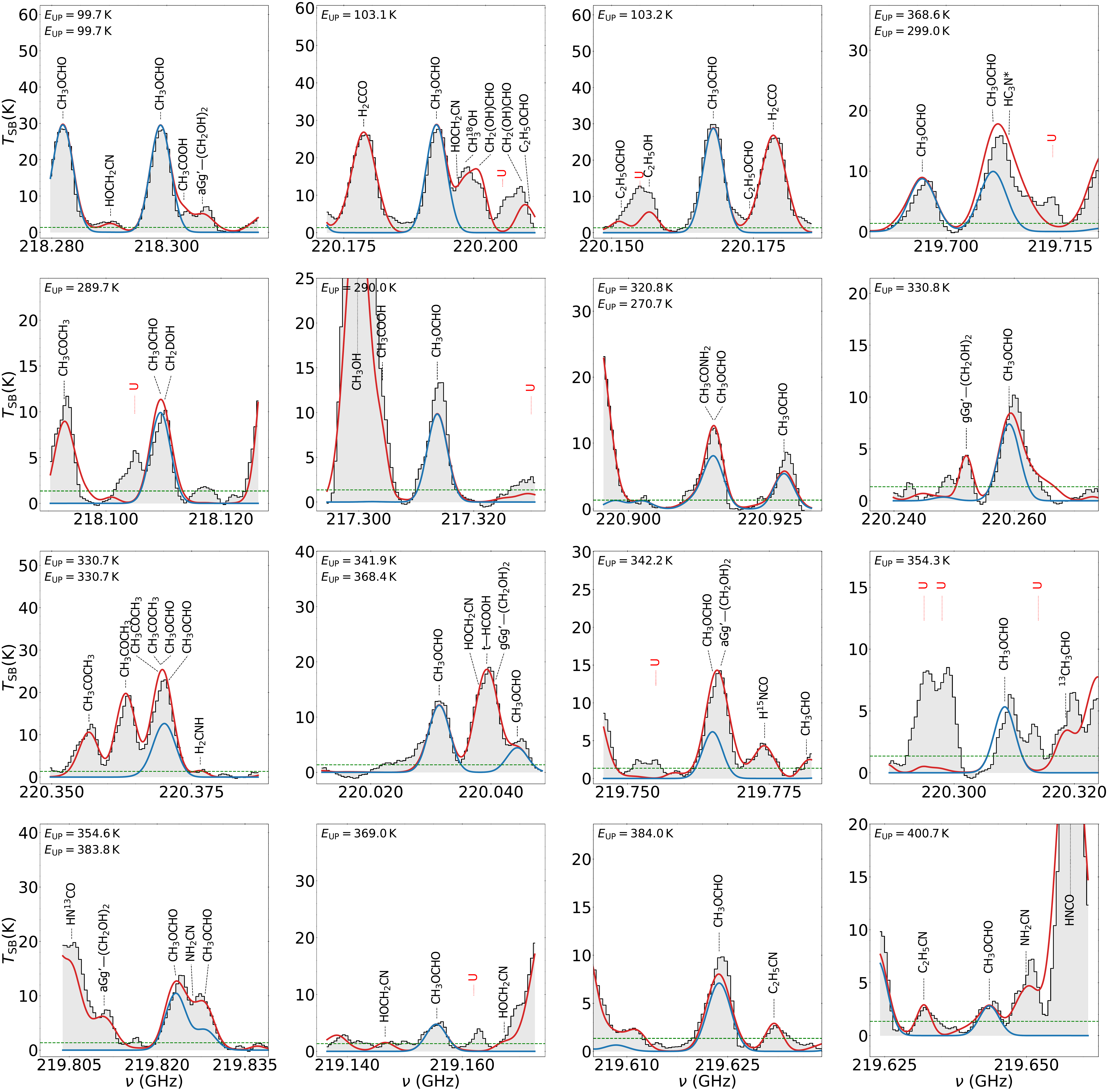}
    \caption{Same as Fig. \ref{fig:ga_transitions}, but for MF, as described in Sect. \ref{subsection:MF}.}
    \label{fig:mf_transitions}
\end{figure}

Methyl formate shows the highest number of clean transitions of the three COMs, with a total of 14 unblended transitions. In Fig.\,\ref{fig:mf_transitions} we show selected lines corresponding to MF. To fit the MF emission, we took a total of 41 transitions into account (see Appendix~\ref{macro}), with $\rm 17_{3,14}\rightarrow 16_{3,13}$ at 218297.89 MHz being the brightest. The same transition also has the largest $\tau = 0.185$, being the transition with highest opacity among all the fitted transitions of EG, GA, and MF. In any case, we can still consider all MF transitions as optically thin. The fitted MF transitions cover a wide energy range $E_{\rm up}$ = 69--318\,cm$^{-1}$ or 100--457\,K.

The estimated excitation temperature, $T_{\rm ex} = 167\pm 3$\,K, is consistent with that derived for EG and GA. The column density is one order of magnitude higher $N=(2.2\pm0.4)\times 10^{17}$\,cm$^{-2}$ than those of EG and GA, which indicates that MF is more abundant, as already observed in other works \citep{hollis_thespatialscale, guapos1}. In the case of MF, we did not fix any parameter. We obtained a value for the $\rm FWHM=5.59 \pm 0.08$\,km\,s$^{-1}$ and $V = -35.22 \pm 0.04$\,km\,s$^{-1}$, which are similar to those of EG and GA.

\subsection{Integrated intensity, $T_{\rm ex}$, and $N$ maps of EG, GA, and MF}
\label{subsect:physical_maps}

\begin{figure*}[h!]
  \centering
    \includegraphics[width=\textwidth]{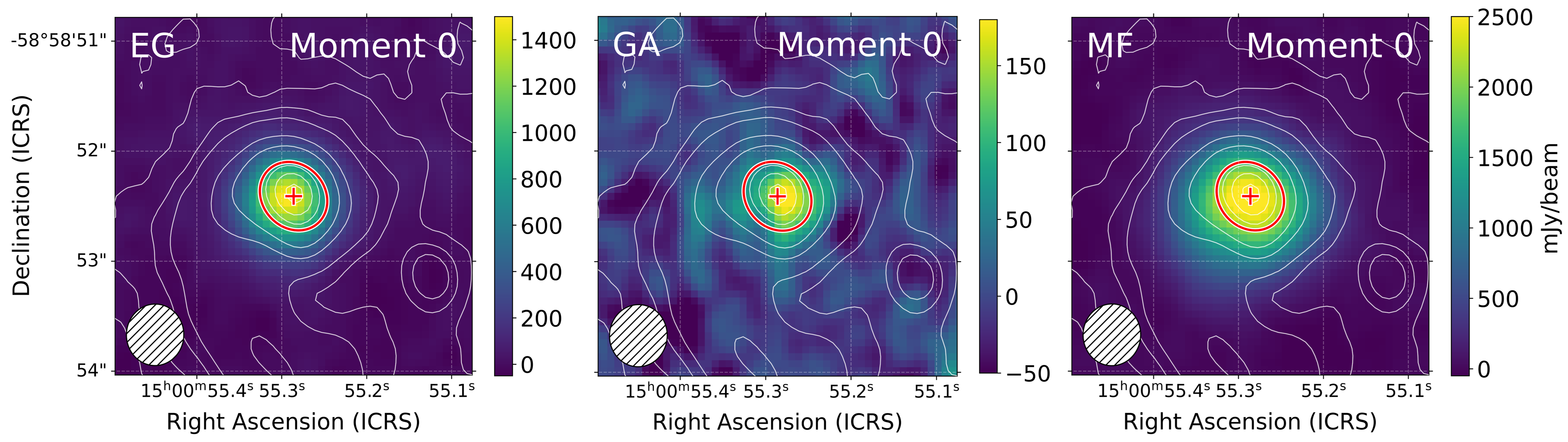}
  \caption{Integrated-intensity maps for EG (left), GA (middle), and MF (right). The white contours correspond to the continuum emission at $5\sigma$, $10\sigma$, $20\sigma$, $30\sigma$, $60\sigma$, $90\sigma$, $180\sigma$, $220\sigma$, and $260\sigma$ with $\sigma=0.28\rm\,mJy\, beam^{-1}$, and the red cross shows its peak position. The red ellipses surrounded by white show the location and size of the dust continuum core AG318$-$c9 as estimated by \cite{almagal3}. The hatched ellipse shown in the bottom left corner represents the synthesised beam of the interferometer.}
  \label{fig:moment0_maps}
\end{figure*}

\begin{figure*}[h]
  \centering
  \includegraphics[width=\textwidth]{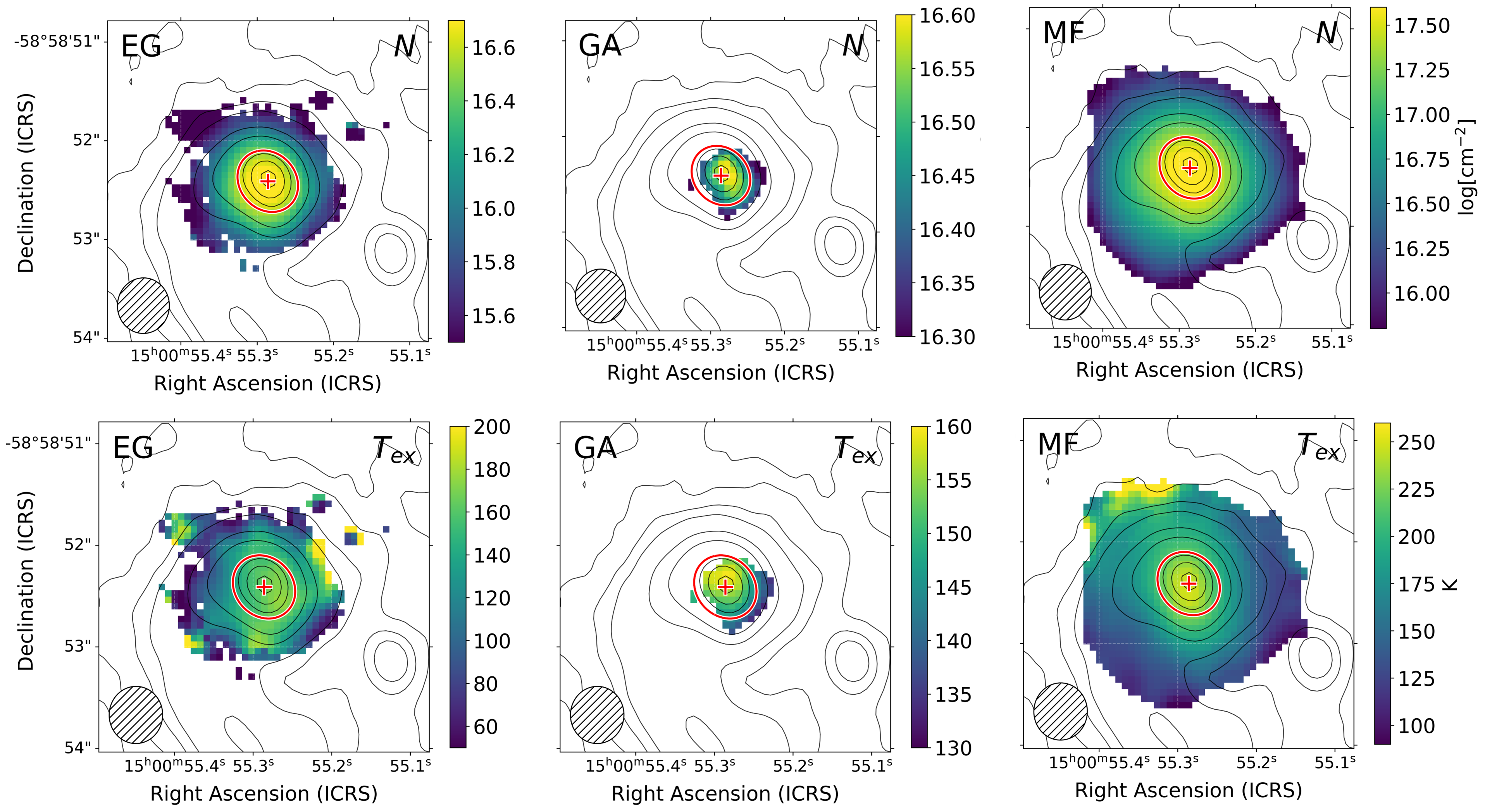}
  \caption{Column density ($N$) maps (top row) and excitation temperature ($T_{\rm ex}$) maps (bottom row) derived for EG (left), GA (middle), and MF (right). The corresponding relative errors are shown in Appendix~\ref{error_maps}. The black contours correspond to the continuum emission at $5\sigma$, $10\sigma$, $20\sigma$, $30\sigma$, $60\sigma$, $90\sigma$, $180\sigma$, $220\sigma$, and $260\sigma$ with $\sigma=0.28\rm\,mJy\, beam^{-1}$, and the red cross shows its peak position. The red ellipses surrounded by white show the location and size of the dust continuum core AG318$-$c9 as estimated by \cite{almagal3}. The hatched ellipse shown in the bottom left corner represents the synthesised beam of the interferometer.}
  \label{fig:maps}
\end{figure*}

\begin{figure}[h]
  \centering
    \includegraphics[width=\columnwidth]{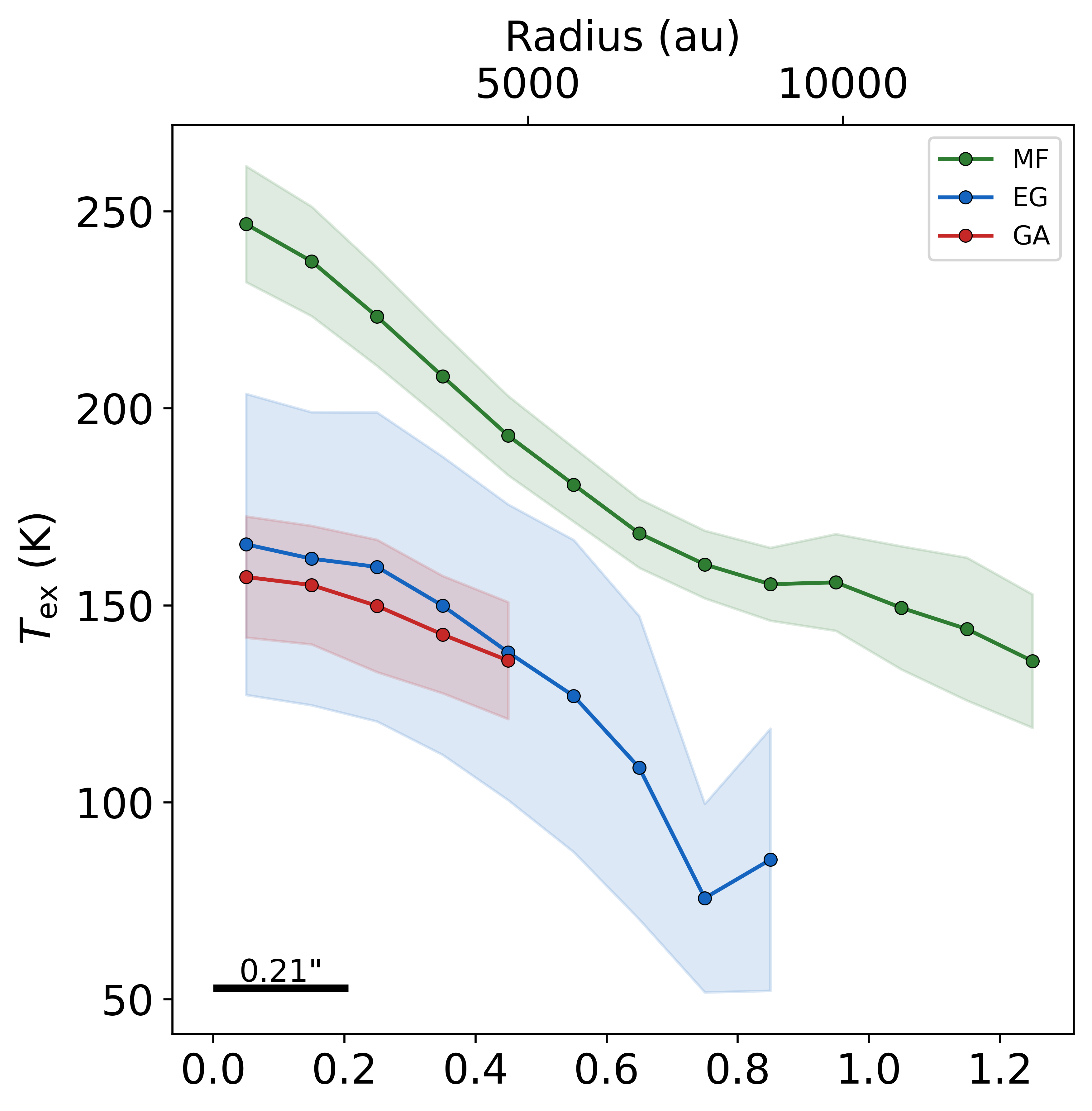} \\
    \includegraphics[width=\columnwidth]{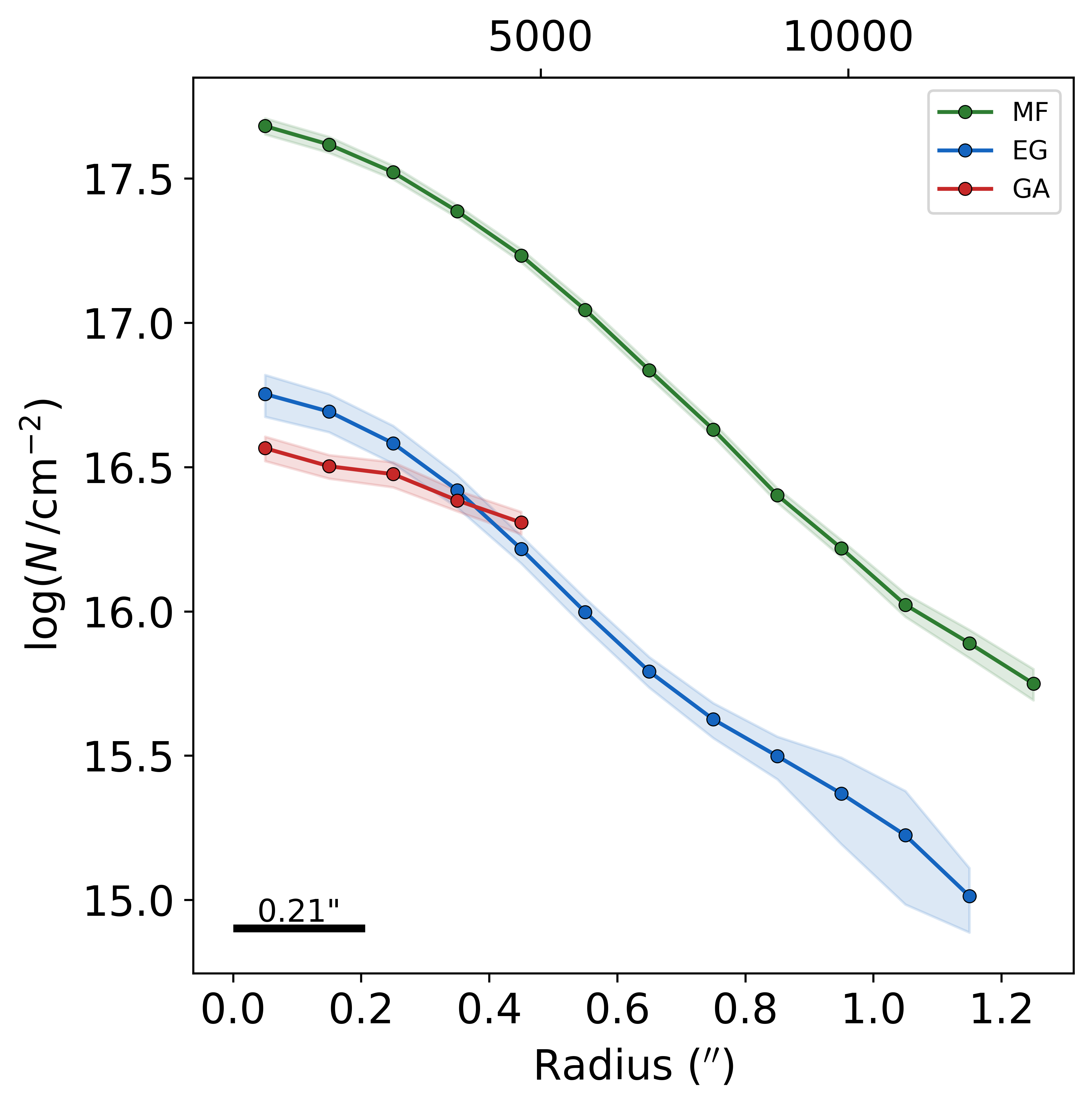} \\
  \caption{Radial profiles derived from the column density ($N$; top) and excitation temperature ($T_{\rm ex}$; bottom) maps. The blue, red, and green curves correspond to EG, GA, and MF, respectively. The shaded areas indicate the measurement uncertainties (see Appendix~\ref{error_maps}). The bar in the lower left corner indicates the radius corresponding to the geometric mean of the synthetic beam.} 
  \label{fig:radial_profile}
\end{figure}

A key aspect of our analysis is to understand how EG, GA, and MF trace the physical environment of AG318$-$c9. To this end, we used MADCUBA to compute their integrated intensity (moment 0), $N$ and $T_{\rm ex}$ maps (Appendix\,\ref{sec:maps}), together with radial profiles of $N$ and $T_{\rm ex}$ obtained by averaging the values within annuli of 0$\farcs$1. The resulting maps and profiles are presented in Figs.~\ref{fig:moment0_maps}--\ref{fig:radial_profile}, and the corresponding relative uncertainty maps are shown in Appendix~\ref{error_maps}.

To generate the maps, we initially attempted to use the same transitions as we adopted to derive the physical parameters presented in Sect.~\ref{sect:results}. However, for EG and GA, a reduced set of transitions was selected, prioritising the least blended transitions while preserving a broad energy coverage. This choice was necessary to ensure convergence of the fits, particularly in the outer core regions with a lower S/N, where the inclusion of blended transitions prevented reliable solutions. The transitions used to produce the maps are listed in Appendix~\ref{transitions_eg_ga_maps}. For EG, the selected transitions cover the same energy range as used in the global fit (E$_{\rm up}$ = 114--580\,K), while for GA, they span a similar range ($E_{\rm up}$ = 37--410\,K).

The moment 0 maps shown in Fig.~\ref{fig:moment0_maps} were constructed using representative transitions for each molecule: the $\rm 35_{10,26}\rightarrow35_{9,27}$ transition at 220.055\,GHz for GA ($E_{\rm up}=410$\,K), the $\rm 24_{1,24}\rightarrow23_{1,23}$ transition at 217.45\,GHz for EG ($E_{\rm up}=117$\,K), and the $\rm 17_{3,14}\rightarrow16_{3,13}$ transition at 218.298\,GHz for MF ($E_{\rm up}=100$\,K). While the EG and MF moment 0 maps were generated with comparable upper-level energies, the GA map relied on a significantly higher $E_{\rm up}$ transition. This difference requires caution when interpreting relative spatial extents based solely on moment 0 emission. Although the moment 0 maps were required to automate the generation of the $N$ and $T_{\rm ex}$ maps, they were not used to derive these quantities directly. Instead, they were used solely to estimate the noise level, which was then applied to assess the reliability of the pixel-by-pixel spectral fits and to exclude solutions below a $3\sigma$ threshold (Appendix~\ref{sec:maps}). The $N$ and $T_{\rm ex}$ maps are therefore based on multi-transition fits and are not biased by the specific energy of the moment 0 transition.

To generate $N$ and $T_{\rm ex}$ maps, only the target species was considered in each case, without considering the contribution of other molecules (Appendix~\ref{sec:maps}). As a result, the core-averaged physical parameters reported in Table~\ref{Table:physical_parameters}, and \ref{Tableapp:physical_parameters}, which account for all detected species, are more robust. Nevertheless, the values derived from the maps are consistent with the core-averaged quantities.

Of the three species, MF is most spatially extended. Its emission closely follows the dust continuum emission and peaks near the continuum maximum (Fig.~\ref{fig:moment0_maps}). The $N(\rm MF)$ map (Fig.~\ref{fig:maps}) reveals a very similar morphology, with a clear column density gradient ranging from $\sim 10^{16}\,\rm cm^{-2}$ in the outer regions to $\sim 3.16\times10^{17}\, \rm cm^{-2}$ near the continuum peak, corresponding to the highest $N$ among the three COMs. The gradient is clearly reflected in the radial profile (Fig.~\ref{fig:radial_profile}), which decreases steadily from the core centre towards larger radii. The $T_{\rm ex}(\rm MF)$ map (Fig.~\ref{fig:maps}) shows a robust temperature gradient, rising from $\sim 100\,$K in the outer parts of the core up to $\sim 250\,$K towards the centre. The radial profile presents a more rapid temperature increase inward of 0$\farcs$8. This combination of large spatial extent and wide $T_{\rm ex}$ dynamic range confirms MF as an excellent tracer of the gas temperature and its gradients in HMCs (e.g. \citealt{Beltran_2018}). These temperatures are consistent with the high rotational temperatures (120--300 K) expected in hot cores \citep{herbst} and typical $T_{\rm ex} \sim 150\,$K estimates \citep{cesaroni_hmc, bisschop2007, rivilla2017}. Because MF has the most unblended transitions, its $T_{\rm ex}$ map has the lowest relative error ($\frac{\Delta T_{\rm ex}}{T_{\rm ex}} \sim 10\,\%$), which increases only in the outer regions with a lower S/N (Appendix~\ref{error_maps}).

The distribution of EG is more compact. Its integrated-intensity and column density maps (Figs.~\ref{fig:moment0_maps} and \ref{fig:maps}) are more concentrated towards the inner core. The $N(\rm EG)$ increases from $\sim 10^{15}\,\rm cm^{-2}$ at the edges to $\sim 5\times10^{16}\,\rm cm^{-2}$ towards the centre. The relative errors for $N(\rm EG)$ increase slightly in the centre of the emission (Fig.~\ref{fig:error_logn}). The $T_{\rm ex}\rm (EG)$ map (Fig.~\ref{fig:maps}) does not exhibit a well-defined peak as clearly as for MF. For clarity, the map is truncated at a radius of 0$\farcs$9 beyond which the uncertainties become too large for a reliable interpretation. Within this radius, the highest $T_{\rm ex}$ values are found within the core, coinciding with the region of strongest dust continuum emission. Overall, the temperature distribution appears to be relatively uniform throughout the inner region where EG emission is detected, with a relative error of $\sim 30\,\%$ (Fig.~\ref{fig:error_tex}). This behaviour might partly reflect the greater difficulty in constraining the fits due to fewer unblended lines, as reflected in the higher relative error, which increases considerably in the outer colder region. Nevertheless, the radial profile clearly decreases in temperature from $\sim 170\,\rm K$ to $\sim 80\,\rm K$ with increasing distance from the core centre, although the profile appears to be less steep in the innermost regions.

The most compact species is GA, with emission confined to the innermost region of the core, close to the continuum peak. The $N(\rm GA)$ and $T_{\rm ex}\rm (GA)$ maps peak centrally, with column densities ranging from $\sim 2\times 10^{16}\,\rm cm^{-2}$ to $\sim 4\times 10^{16}\,\rm cm^{-2}$ and an excitation temperature between $\sim 130\,\rm K$ to $\sim 160\,\rm K$. The two map fits have a relative error of $\sim 10-15\,\%$ (Appendix~\ref{error_maps}). Over the radial range in which EG and GA are detected, the $N$ gradient is comparable to that estimated in EG.

To evaluate whether the different spatial extents of MF, EG, and GA are driven by sensitivity limitations or reflect intrinsic chemical differentiation, we estimated the column density corresponding to a $5\sigma$ detection threshold ($\sim 1.35\,$K) for each molecule with MADCUBA. For MF and EG, the $5\sigma$ limits ($N\sim 10^{16}\,\rm cm^{-2}$ and $N\sim 2.5\times 10^{15}\,\rm cm^{-2}$, respectively) closely match the minimum column densities reached by their radial profiles, indicating that the apparent truncation is largely driven by sensitivity. In contrast, the $5\sigma$ threshold for GA is $N(\rm GA)\approx 2\times 10^{15}\,\rm cm^{-2}$, whereas the GA radial profile only extends to $N(\rm GA)\approx 1.6\times 10^{16}\,\rm cm^{-2}$, almost one order of magnitude. This might suggest that the lack of detectable GA emission beyond the innermost regions of the core is not limited by sensitivity, but instead reflects its intrinsically compact distribution.

The compact emission of EG and GA suggests that they are excellent candidates for tracing the innermost high-density regions of the core. Due to their optically thin transitions, they might also be used to probe the kinematics within these regions. Furthermore, the fact that EG and GA are only detected in the warm inner region, unlike the more extended MF, might imply that a specific temperature threshold must be reached to sublimate them from the icy dust grains. This observation strongly supports a shared grain-surface formation pathway, consistent with models of dust-grain chemistry (e.g. \citealt{fedoseev2015, Simons2020, garrod2022}).

However, the innermost region must be interpreted with caution due to observational effects. At radii comparable to or smaller than the beam size ($\sim 0\farcs21$), the observed brightness temperature might be affected by beam dilution. In addition, dust opacity might become non-negligible towards the core centre, potentially attenuating line emission from the deepest layers. These effects might result in an underestimation of the true column densities and excitation temperatures in the central region, particularly for compact species such as EG and GA. Moreover, as shown in Fig.~\ref{fig:radial_profile}, the emission of the three species extends beyond the beam size, and therefore, it is expected to be hardly affected by beam dilution.

The observed spatial differentiation among the three species suggests that their distributions are not governed solely by thermal desorption, but also by their formation pathways and ice chemistry. MF, with a lower binding energy ($\sim 4000$\,K; desorption at $\sim 100-120$\,K), is expected to be more extended and associated with CO-rich outer regions, whereas GA has a higher binding energy ($\sim 5900$\,K; $\sim 140-190$\,K) and is therefore expected to trace warmer and denser regions, consistent with its more compact emission \citep{oberg_2009}. EG, with an even higher binding energy ($\sim 7500$\,K; $\sim 150-180$\,K), does not show a correspondingly more compact distribution, suggesting that its spatial extent cannot be explained by desorption alone. Instead, its distribution might reflect efficient formation during the warm-up phase, such as via CH$_2$OH recombination and chemical stability once released into the gas phase \citep{oberg_2009, fedoseev2015}. Overall, this chemical differentiation supports a scenario in which the temperature gradients and the initial ice composition can play a key role in shaping the spatial distribution of COMs \citep{garrod_2008}.

\section{Conclusions}

We presented a comprehensive analysis of the chemical reservoir of the AG318.9477$-$0.1960 clump core 9 using data from the ALMAGAL project and a frequency range of $\sim 217-221\,\rm GHz$. Throughout the $\sim\, 4\rm\, GHz$ setup, we identified 65 species, of which 44 are O-bearing species, 28 are N-bearing, 8 are S-bearing, and 2 are Si-bearing. Thirty-three of these species are COMs.  The fact that most of the detections are O-bearing molecules suggests an efficient thermal sublimation of the ice mantles into the gas phase. For all the detected species, we derived their physical parameters and abundances with respect to H$_2$. 
 
This spectral analysis allowed us to characterise weak and blended species, such as the three COMs of interest: EG, GA, and MF. For EG, we covered an $E_{\rm UP}$ range of $114-580$\,K and detected 8 unblended transitions. We found that the species that are more strongly blended with EG are MF, $\rm CH_3NCO$, $\rm HNCO$, $\rm HN^{13}CO$, $\rm C_2H_5OH$, and $\rm CHD_2OH$. For GA, we traced transitions with $E_{\rm UP}=37-450$\,K including 3 unblended lines. The main blending species are $\rm CH_3COCH_3$, $\rm CH_3^{18}OH$, $\rm CH_3OCH_3$, and $\rm CH_3NCO$. Finally, MF exhibited transitions with $E_{\rm UP}=100-457$\,K and 14 unblended lines, making it the species least affected by blending. All three species trace warm and dense gas, although they probe distinct thermal components. EG traces the warmest material, with a $T_{\rm ex}$ of $\sim 240\,\rm K$. GA and MF trace slightly cooler gas, with $T_{\rm ex}$ $\sim 168\,\rm K$ and $\rm \sim 167\,K$, respectively. Regarding $N$ and abundance, MF has a column density of $\sim2\times10^{17}$\,cm$^{-2}$ and is clearly the most abundant species, with $X\sim 3\times10^{-8}$. This abundance is significantly higher than that of EG and GA, which have abundances of $\rm \sim 7\times 10^{-9}$ and $\rm \sim 2\times10^{-9}$, and $N$ of $\sim5\times10^{16}$\,cm$^{-2}$ and $\sim1.5\times10^{16}$\,cm$^{-2}$, respectively.

The chemical comparison with the chemically rich HMC G31.41+0.31 indicated that this core shows systematically higher column densities than AG318$-$c9 for O- and N-bearing COMs. This suggests that G31 is either intrinsically warmer or at a more advanced evolutionary stage than AG318$-$c9, having experienced more efficient thermal sublimation of its ice mantles. This comparison also revealed a chemical differentiation between these families, with a stronger enhancement of O-bearing species in G31, while N-bearing species showed a more moderate contrast between the two sources. Conversely, S-bearing species are more comparable between the two sources. Notably, AG318$-$c9 shows a significant overabundance of the prebiotic molecules $\rm NH_2CN$ and $\rm H_2CNH$, which indicates specific physical conditions or precursor availability for its formation.

Our analysis of the $T_{\rm ex}$ and $N$ maps for EG, GA, and MF within AG318$-$c9 revealed distinct morphologies. The emission of the three species extends beyond the beam size and is therefore not significantly affected by beam dilution. However, observational effects, in particular, dust opacity, might still lead to underestimated column densities and excitation temperatures in the innermost region. MF shows significantly higher column densities ($N \approx 3\times 10^{17}\, \rm cm^{-2}$) and a spatially extended distribution, robustly tracing the thermal structure of the core over a broad temperature range ($T_{\rm ex} \sim 100-250\,$K). In sharp contrast, EG and GA are compact and confined to the innermost core region, with GA being particularly compact. They both exhibit similar lower column densities ($N \approx 4\times10^{16}\,\rm cm^{-2}$). Their restricted distribution indicates that they only trace the warmest densest gas, implying a high sublimation temperature threshold. This might support a shared grain-surface formation pathway. 

As a methodological output of this work, we developed two MADCUBA scripts that will be used for future molecular analysis studies within the ALMAGAL survey. Work is now underway on a complete study of EG, GA, and MF in all ALMAGAL cores to understand the correlations between these COMs, the physical conditions that lead to their formation, and their possible chemical pathways.

\section*{Data availability}

The scripts used for the automated line identification and map generation, as well as the complete table of molecular transitions used for the MADCUBA fits of the molecular emission towards AG318$-$c9 (see Tables~\ref{tab:detected_species}, \ref{Table:physical_parameters}, and \ref{Tableapp:physical_parameters}) are available at: \url{https://github.com/JofreAllande/ALMAGAL/}.

\begin{acknowledgements}
We thank the anonymous referee for the useful comments.

The table of molecular transitions were made thanks to the Python library \texttt{RichValues}\footnote{\url{https://github.com/andresmegias/richvalues}} created by Andr\'es Meg\'ias.

J.A. acknowledges the SKA PhD fellowship "Astrochemistry of prebiotic molecules in high-mass star-forming regions in preparation of the SKA observations" offered by INAF. C.Y.L. acknowledges financial support through the INAF Large Grant ``The role of MAGnetic fields
in MAssive star formation'' (MAGMA).

R.S.K acknowledges financial support from the ERC via Synergy Grant ``ECOGAL'' (project ID 855130) and from the German Excellence Strategy via the Heidelberg Cluster ``STRUCTURES'' (EXC 2181 - 390900948). In addition R.S.K. is grateful for funding from the German Ministry for Economic Affairs and Climate Action in project ``MAINN'' (funding ID 50OO2206), and from DFG and ANR for project ``STARCLUSTERS'' (funding ID KL 1358/22-1). 

L.C. acknowledges support from the grant PID2022-136814NB-I00 by the Spanish Ministry of Science, Innovation and Universities/State Agency of Research MICIU/AEI/10.13039/501100011033 and by ERDF, UE. The project that gave rise to these results received the support of a fellowship from the “la Caixa” Foundation (ID 100010434). The fellowship code is LCF/BQ/PR25/12110012.

L.B. gratefully acknowledges support by the ANID BASAL project FB210003.

C.M. and the INAF-IAPS team acknowledge funding from the European Research Council (ERC) under the European Union’s Horizon 2020 program through the ECOGAL Synergy grant (ID 855130). C.M. acknowledges funding from INAF Mini Grants RSN2 2024 "Zodyac" CUP C83C25000340005.

R.K. acknowledges financial support via the Heisenberg Research Grant funded by the Deutsche Forschungsgemeinschaft (DFG, German Research Foundation) under grant no.~KU 2849/9, project no.~445783058.

A.S.-M.\ acknowledges support from the RyC2021-032892-I grant funded by MCIN/AEI/10.13039/501100011033 and by the European Union `Next GenerationEU’/PRTR, as well as the program Unidad de Excelencia María de Maeztu CEX2020-001058-M, and support from the PID2020-117710GB-I00 (MCI-AEI-FEDER, UE).

A. L-G. acknowledges support from the grant PID2022-136814NB-I00 by the Spanish Ministry of Science, Innovation and Universities/State Agency of Research MICIU/AEI/10.13039/501100011033 and by ERDF, UE; and from the Consejo Superior de Investigaciones Cient{\'i}ficas (CSIC) and the Centro de Astrobiolog{\'i}a (CAB) through the project 20225AT015 (Proyectos intramurales especiales del CSIC).

The National Radio Astronomy Observatory is a facility of the National Science Foundation operated under cooperative agreement by Associated Universities, Inc.

Part of this research was carried out at the Jet Propulsion Laboratory, California Institute of Technology, under a contract with the National Aeronautics and Space Administration (80NM0018D0004).

\end{acknowledgements}

\bibliography{bibliografia}

\begin{appendix}

\section{Automatization with MADCUBA}
\label{macro}

We developed a MADCUBA macro to automate line identification across the ALMAGAL sample. Molecular species identified toward AG318$-$c9 using the detailed line identification described in Sect.~\ref{subsect: line identification} were added to the SLIM module using the CDMS\footnote{\url{https://cdms.astro.uni-koeln.de/classic/}} \citep{cdms2001,cdms2005, cdms2016} or the JPL\footnote{\url{https://spec.jpl.nasa.gov/ftp/pub/catalog/catdir.html}} \citep{jpl} catalogues. As previously mentioned, for each species only sufficiently strong ($\rm S/N\geq5$) and partially unblended transitions were selected, ensuring that AUTOFIT operates on high-quality lines and minimising potential errors from blended or contaminated transitions. In future studies where the entire ALMAGAL sample will be analysed and chemically poorer sources will be included, the presence of species detected in AG318$-$c9 will not pose a problem, since AUTOFIT will not converge for species absent in the observed spectra and will not affect the fitting of other species. 

\begin{figure}[h]
    \centering
    \includegraphics[width=0.95\columnwidth]{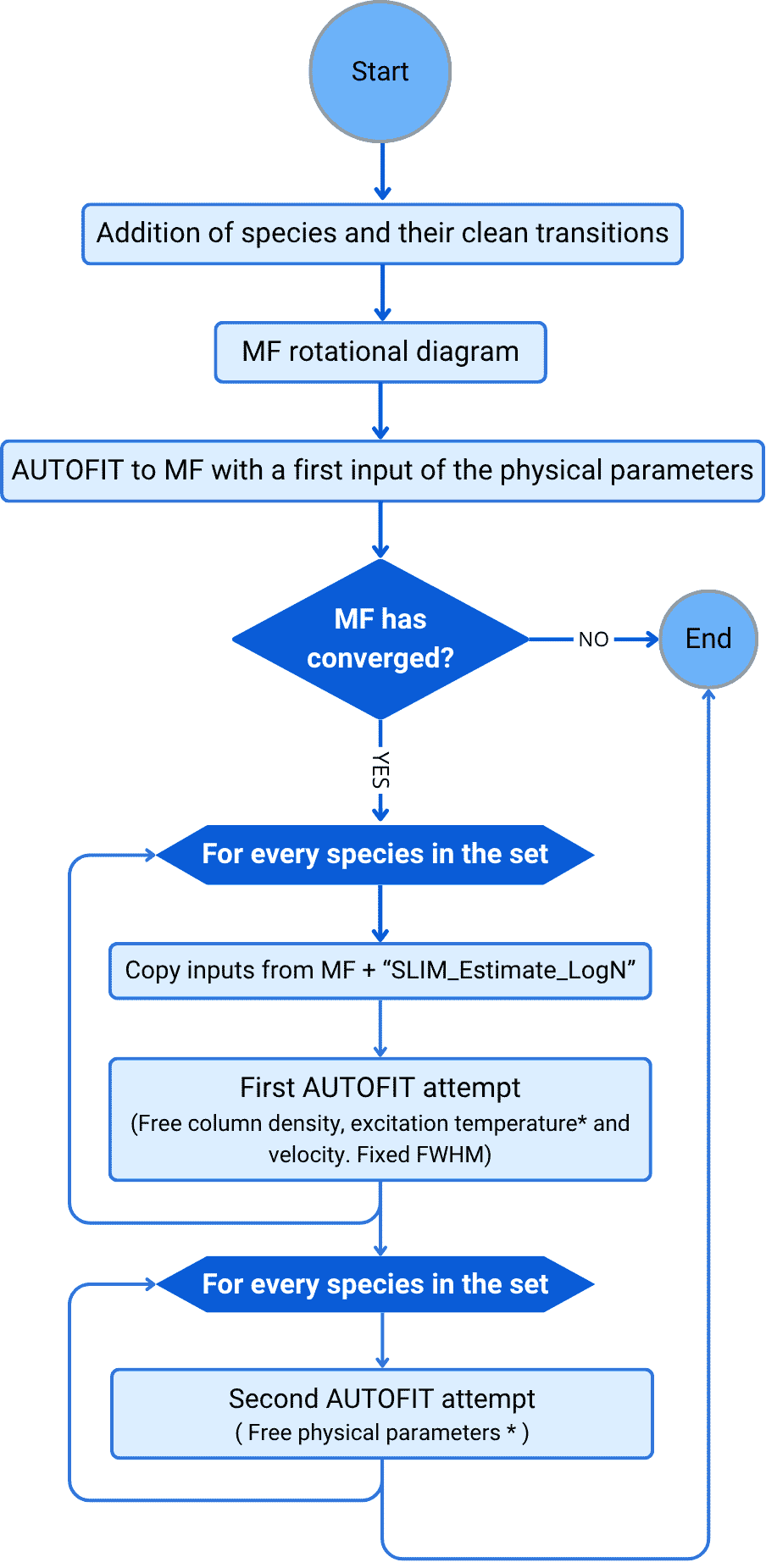}
    \caption{Flowchart of the MADCUBA macro that automates the line identification process. The asterisk (*) indicates that physical parameters are kept free whenever possible, following the criteria described in the text.}
    \label{fig:flowchart_macro}
\end{figure}

SLIM requires an initial input of $N$, $T_{\rm ex}$, $V$, FWHM, and $\theta_s$ for each species to run AUTOFIT. If these initial values are absent or far from realistic estimates, the programme does not converge. For AG318$-$c9, we assumed as initial guesses $V$ = $-$34.4\,km\,s$^{-1}$, which is the systemic velocity of the clump, and FWHM = 5\,km\,s$^{-1}$, which is the typical width of the lines for this source, as obtained from simple Gaussian fits to some species in our setup. The parameter $\theta_s$ was fixed to 0, which in MADCUBA means that the source emission size is equal to or larger than the synthesised beam and hence unaffected by beam dilution. This assumption is appropriate for AG318$-$c9. To reliably estimate $T_{\rm ex}$, we require multiple and unblended transitions spanning a wide energy range. Because this is not possible for all the species, particularly if a species has a single transition, we decided to use a common $T_{\rm ex}$ as an initial input, estimated from a species with many transitions covering a broad range of energies. To achieve this automatically, we have estimated this common $T_{\rm ex}$ by calculating the rotational diagram of this species with MADCUBA. Although we initially wanted to use \MCN\ to estimate such a common initial $T_{\rm ex}$, we realised that the lowest $K$ transitions were saturated, indicating that they are optically thick, leading to an overestimate of the rotational temperature ($T_{\rm rot}$) and an underestimate of the rotational column density ($N_{\rm rot}$). Therefore, we used a total of 41 selected optically thin MF transitions that contribute significantly to the observed spectrum. Among these, 14 are non-blended and 27 are slightly blended. The selected transitions offer a broad coverage of energies ($\rm E_{\rm up}$ 99.72\,K - 429.4\,K or $\rm 69.31\,cm^{-1}\,-\,298.45\,cm^{-1}$) that allowed us to properly calculate the rotational diagram of MF and obtain $T_{\rm rot}$, 
which has been used as initial seed for $T_{\rm ex}$ for all species. The rotational column density obtained from the diagram has also been used as initial guess for $N$ to fit MF. 

If AUTOFIT converges for MF with these initial parameters, the same $T_{\rm ex}$, $V$, and FWHM values are applied, as initial guesses, for the remaining species, with the exception of $N$. Assuming starting common values of $T_{\rm ex}$, $V$ and FWHM for different species tracing the HMC is reasonable, as they are not expected to vary significantly. In contrast, $N$ can change significantly and is the parameter for which the synthetic spectrum is most sensitive. To address this, SLIM provides the \texttt{"SLIM\_Estimate\_LogN"}\footnote{\url{https://cab.inta-csic.es/madcuba/documentation.html}}, that estimates $\log(N)$ for a given species by using the transition with the highest expected intensity based on the molecular parameters ($T_{\rm ex}$, V, and FWHM). This estimate is derived from the velocity-integrated intensity obtained by integrating the line emission in the velocity range defined by the linewidth. 

Once appropriate inputs are obtained for all the species, a first AUTOFIT is performed in two rounds. In the first round, the FWHM remains fixed, and each species is fitted individually without taking into account the contribution from other species (disabling the ALL SPECIES flag in AUTOFIT). During this first step, brighter and unblended molecules converge easily and independently. The converged molecules provide constraints that will help in the second AUTOFIT iteration, which will try to fit all the species, including those blended.  In this second round, the FWHM is left free and the contribution of all blended species is properly taken into account (enabling the ALL SPECIES flag in AUTOFIT).

We have tried to leave as many free physical parameters as possible in order to have the most unforced solution. However, for species with limited transitions, poor energy coverage, or blended lines, we adopted a hierarchical approach by fixing parameters when AUTOFIT does not converge: i) first FWHM, because usually this parameter can easily be obtained from the observed spectra; ii) $V$ is fixed next, using the systemic velocity (V$_0$) as a reference to constrain velocity shifts, since large deviations from this value are not expected; and iii) $T_{\rm ex}$ is fixed when only one transition is available, as for example for H$_2^{13}$CO or C$^{18}$O, or when the number of transitions with different enough energies is not sufficient to make the fit to converge. In our approach, we never fixed $N$. Although this procedure is largely automated, the resulting fit must be visually inspected to ensure consistency with the observations, and this last step should be done manually. For a flowchart of the process, see Fig. \ref{fig:flowchart_macro}. Note that if MF does not converge, the automated macro stops. This could be due to MF not being present in the spectra, being too weak to be properly fit, or due to incorrect initial guesses. In the latter case, we should investigate the reason and, if possible, adjust the initial guesses. 

\section{Generation of the maps ($N$)}
\label{sec:maps}
To generate the $T_{\rm ex}$ and $N$ maps we used the SLIM module to fit the spectra on a pixel-by-pixel basis. MADCUBA provides a built-in tool to generate the integrated intensity map (moment 0), velocity field (moment 1), and velocity dispersion (moment 2) maps for selected transitions. To derive the moment maps, we selected an unblended transition for each of the three molecules of interest. For GA the transition $\rm 35_{10,26} \rightarrow 35_{9,27}$ at 220.055\,GHz, for EG the transition $\rm 24_{1,24} \rightarrow 23_{1,23}$ at 217.45\,GHz, and for MF the transition $\rm 17_{3,14} \rightarrow 16_{3,13}$ at 218.298\,GHz.

To generate the $T_{\rm ex}$ and $N$ maps, we are using the spectral cube (spw0 and spw1) within a user-defined region of interest. The procedure performs the analysis only on pixels with a S/N above a given threshold in the moment 0 map, which in our case was $\rm S/N > 3$. 
A preliminary rotational diagram analysis, for each pixel, is used to estimate the initial guesses for $T_{\rm ex}$ and $\log N$, as detailed in Sect.~\ref{macro}. The velocity ($V$) and linewidth (FWHM) parameters are extracted from the line velocity (moment 1) and velocity dispersion (moment 2) maps, respectively. Moment maps were generated with a built-in tool from MADCUBA. The AUTOFIT routine inside SLIM is then executed for each pixel to perform a full LTE fit. The MADCUBA procedure generates as output a SLIM product containing the individual pixel-by-pixel fits, and the $\log(N)$, $T_{\rm ex}$, $V$, and FWHM maps. The only difference with the macro defined in Sect.~\ref{macro} is that this procedure only takes into account the molecule that we are fitting, and not the contribution of all the other molecules. The provided maps have two channels: channel 1 contains the fitted values, while channel 2 stores the associated uncertainties.

\subsection{Uncertainty $N$ and $T_{\rm ex}$ maps}
\label{error_maps}

\begin{figure}[!h]
  \centering
    \includegraphics[width=\columnwidth]{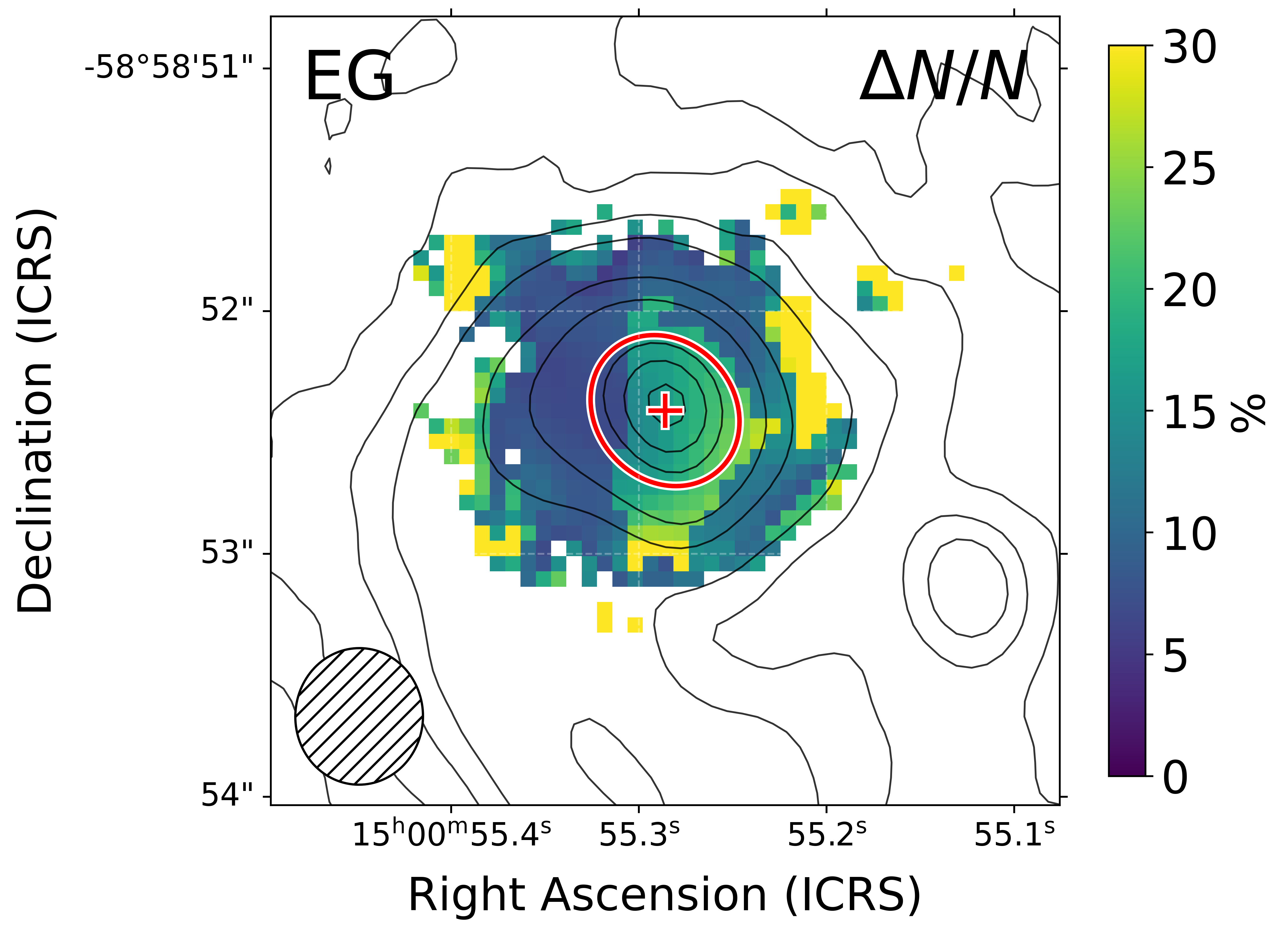}\\
    \includegraphics[width=\columnwidth]{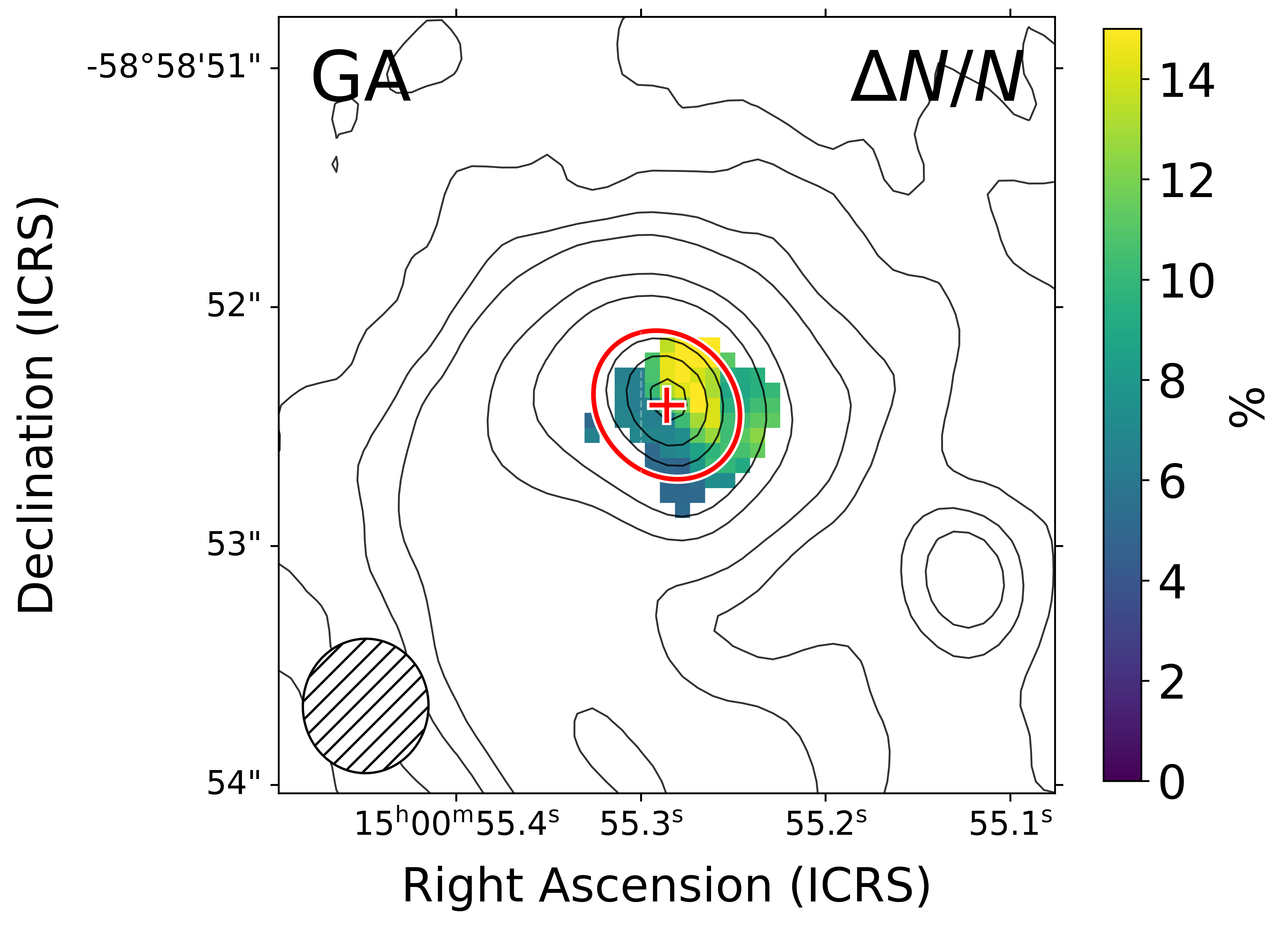}\\
    \includegraphics[width=\columnwidth]{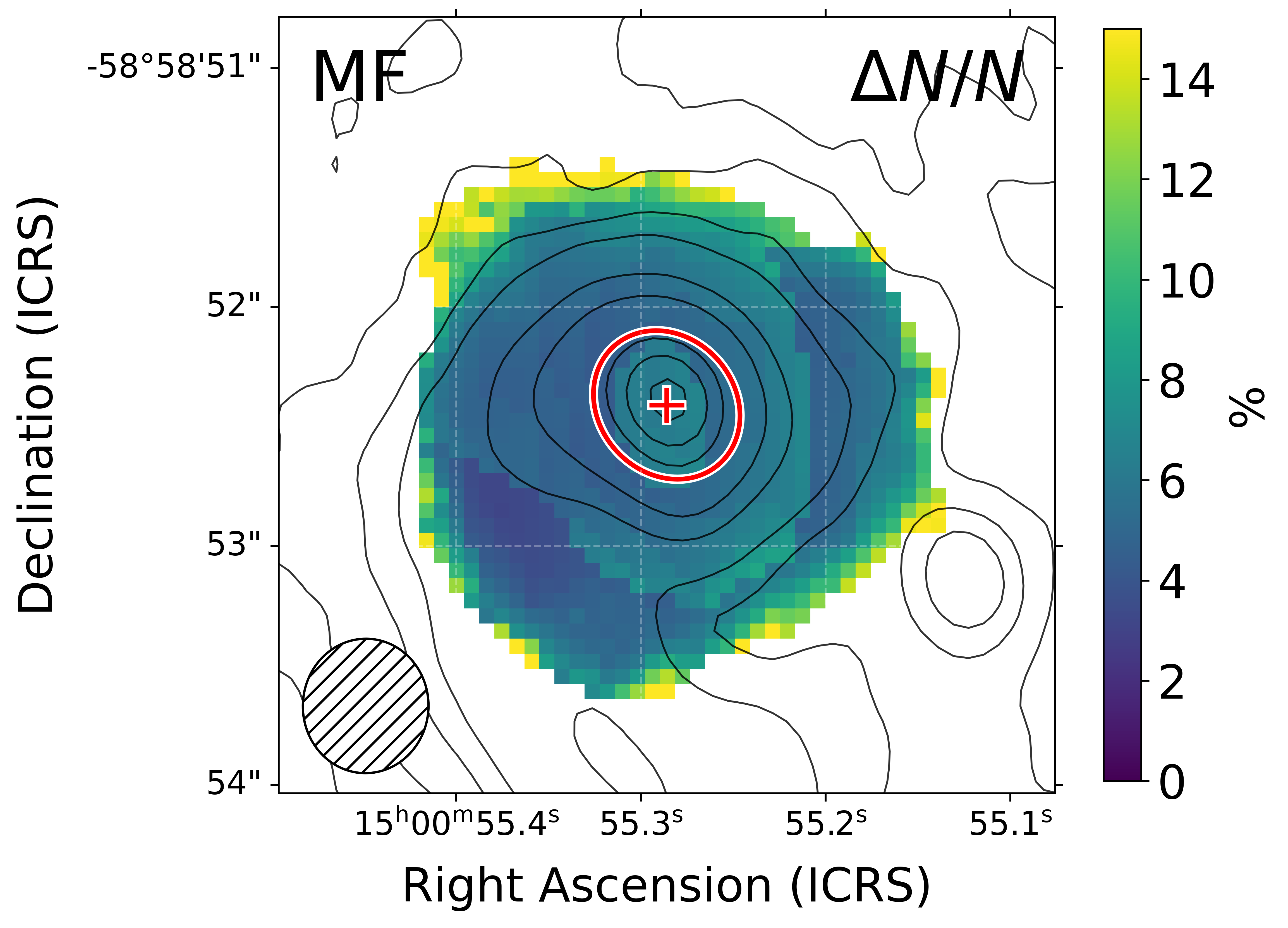}
  \caption{Relative error from the column density ($N$) maps derived for EG (top), GA (middle), and MF (bottom), respectively. Black contours correspond to the continuum emission at $5\sigma$, $10\sigma$, $20\sigma$, $30\sigma$, $60\sigma$, $90\sigma$, $180\sigma$, $220\sigma$, and $260\sigma$ with $\sigma=0.28\rm\,mJy\, beam^{-1}$. The red cross indicates the continuum peak position. The red ellipses outlined in white show the location and size of the dust continuum core AG318$-$c9 as estimated by \cite{almagal3}. The hatched ellipse shown on the bottom left corner represents the synthesised beam of the interferometer.}
  \label{fig:error_logn}
\end{figure}

\begin{figure}[!h]
  \centering
    \includegraphics[width=\columnwidth]{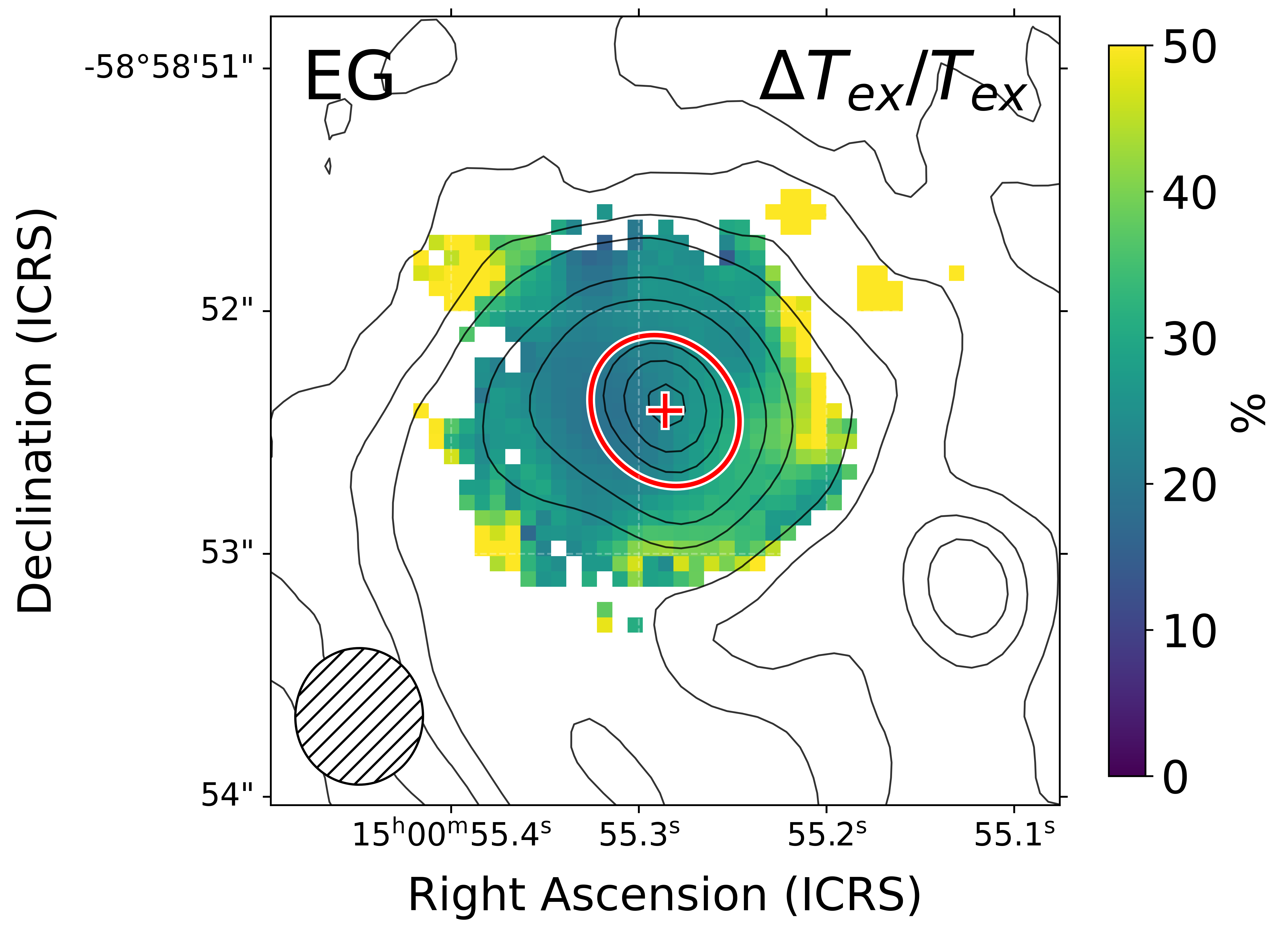}\\
    \vspace{0.1cm}
    \includegraphics[width=\columnwidth]{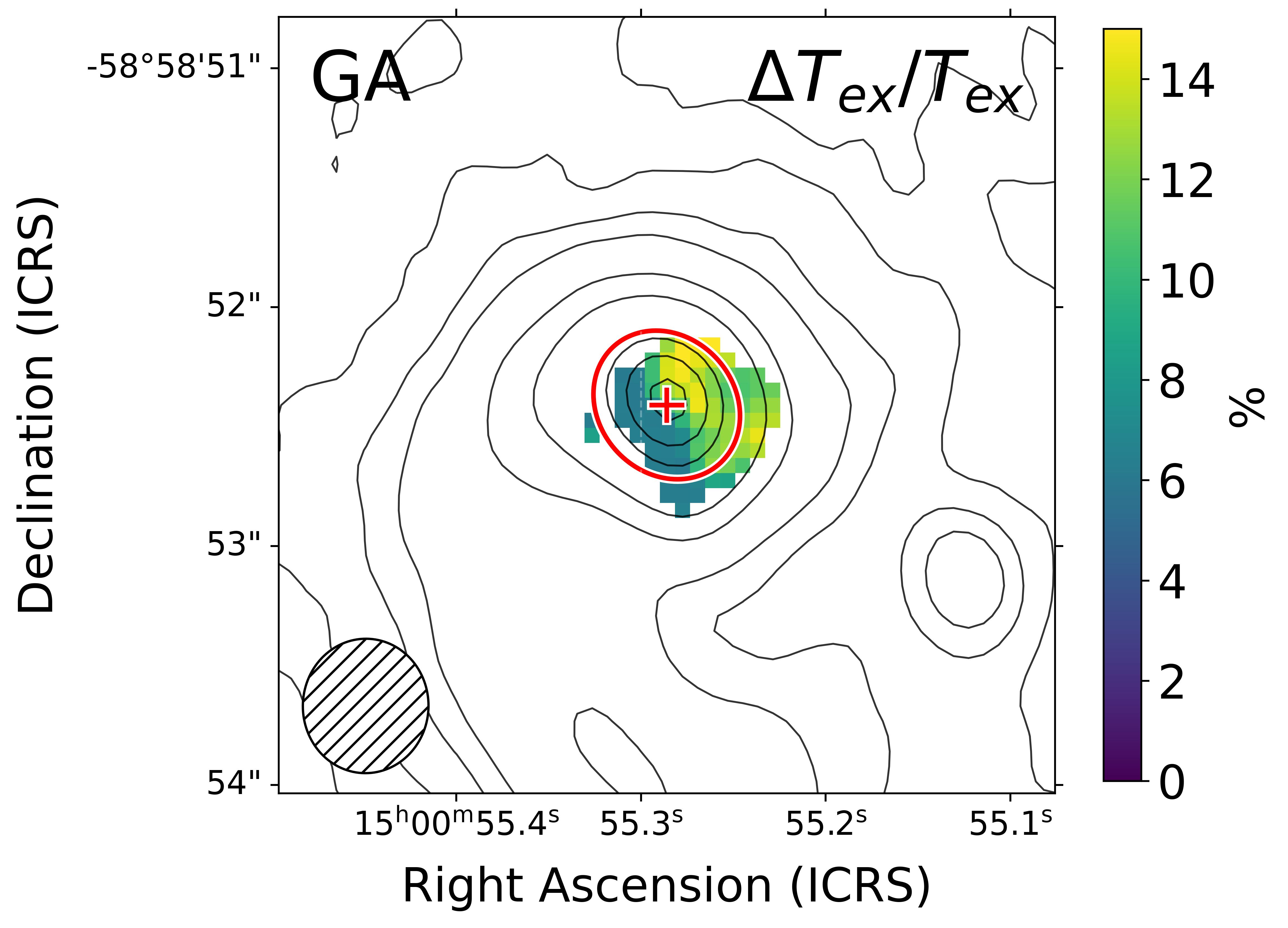}\\
    \vspace{0.1cm}
    \includegraphics[width=\columnwidth]{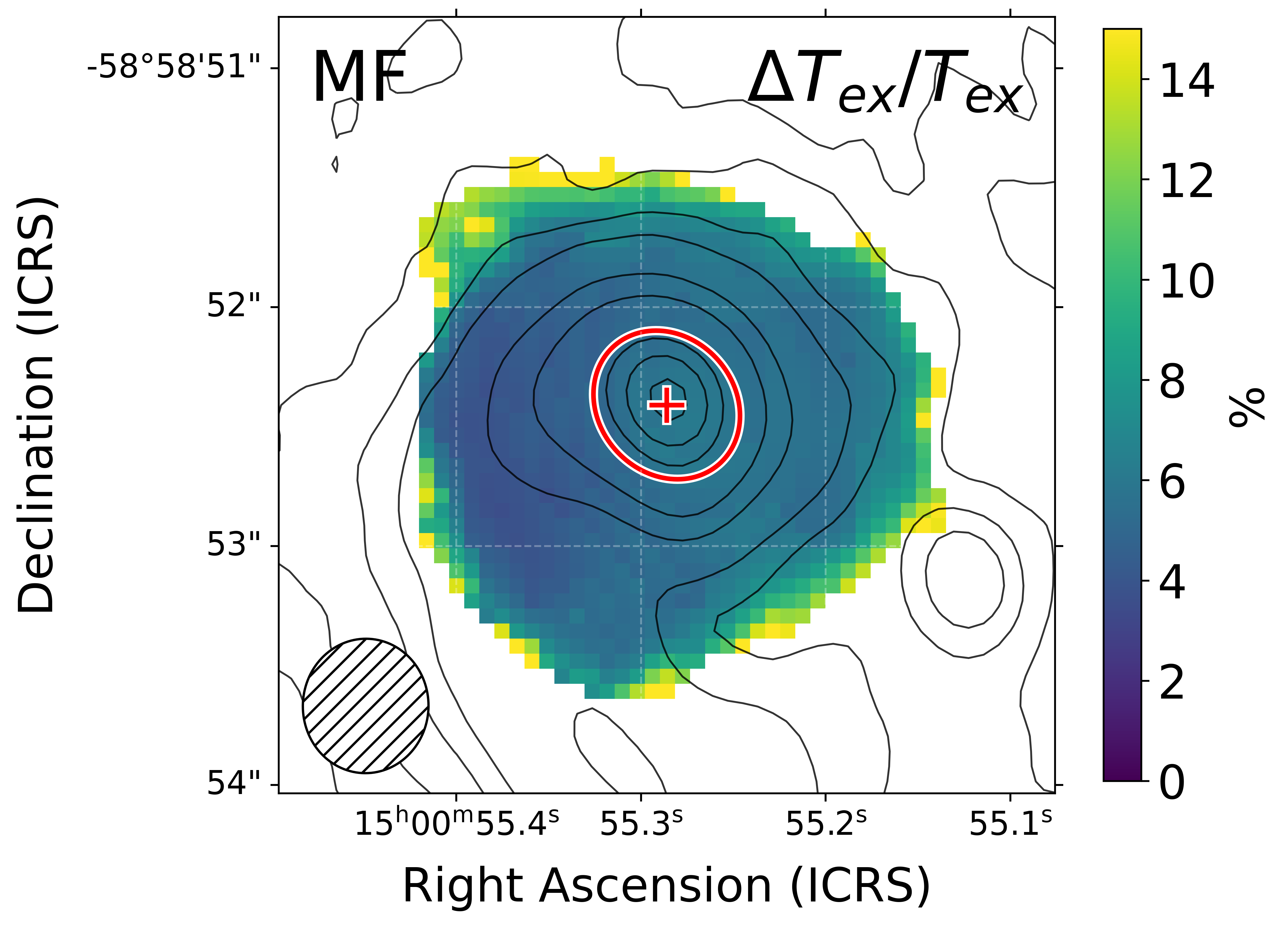}
  \caption{Relative error from excitation temperature ($T_{\rm ex}$) maps derived for EG (top), GA (middle), and MF (bottom), respectively. Black contours correspond to the continuum emission at $5\sigma$, $10\sigma$, $20\sigma$, $30\sigma$, $60\sigma$, $90\sigma$, $180\sigma$, $220\sigma$ and $260\sigma$ with $\sigma=0.28\rm\,mJy\, beam^{-1}$. The red cross indicates the continuum peak position. The red ellipses outlined in white show the location and size of the dust continuum core AG318$-$c9 as estimated by \cite{almagal3}. The hatched ellipse shown on the bottom left corner represents the synthesised beam of the interferometer.}  
  \label{fig:error_tex}
\end{figure}

We present the relative uncertainty maps of the column density ($N$; Fig.~\ref{fig:error_logn}) and excitation temperature ($T_{\rm ex}$; Fig.~\ref{fig:error_tex}) maps derived from the spectral fitting, in order to assess the reliability of the obtained physical parameters. 

\subsection{Molecular transitions used for $N$ and $T_{\rm ex}$ maps for ethylene glycol and glycolaldehyde}
\label{transitions_eg_ga_maps}
We present the molecular transitions used in the MADCUBA pixel-by-pixel fits for EG and GA.

    \begin{table*}[!h]
    \centering
    \caption{Sorted list by frequency of the transitions used in the MADCUBA pixel-by-pixel method (Fig.~\ref{fig:maps}).} 
    \renewcommand*{\arraystretch}{1.2} 
    \begin{tabular}{cccccc} 
        \hline 
        {Frequency\,(GHz)} & {Species} & {Transition} & {log$I$\,($\rm nm^{2}\,MHz$)} & {\bf $E_{\rm UP}\,(\rm K)$} & {\bf $\tau (\times 10^{-2})$} \tabularnewline 
        \hline 
        $217.272$ & CH$_2$(OH)CHO & $31_{4,27,0} \rightarrow 31_{3,28,0}$ & $-3.7$ & $290$ & $1.3 \pm 1.4$ \tabularnewline 
        $217.604$ & CH$_2$(OH)CHO & $24_{5,19,0} \rightarrow 23_{6,18,0}$ & $-3.9$ & $187$ & $1.1 \pm 1.4$ \tabularnewline 
        $217.626$ & CH$_2$(OH)CHO & $9_{5,5,0} \rightarrow 8_{4,4,0}$ & $-3.7$ & $40$ & $3.3 \pm 1.5$ \tabularnewline 
        $217.831$ & CH$_2$(OH)CHO & $9_{5,4,0} \rightarrow 8_{4,5,0}$ & $-3.7$ & $40$ & $3.3 \pm 1.5$ \tabularnewline 
        $218.261$ & CH$_2$(OH)CHO & $20_{3,17,0} \rightarrow 19_{4,16,0}$ & $-3.6$ & $126$ & $3.0 \pm 1.5$ \tabularnewline 
        $219.123$ & CH$_2$(OH)CHO & $29_{3,26,0} \rightarrow 29_{2,27,0}$ & $-3.7$ & $250$ & $1.3 \pm 1.4$ \tabularnewline 
        $220.055$ & CH$_2$(OH)CHO & $35_{10,26,0} \rightarrow 35_{9,27,0}$ & $-3.6$ & $410$ & $0.9 \pm 1.4$ \tabularnewline 
        $220.464$ & CH$_2$(OH)CHO & $20_{2,18,0} \rightarrow 19_{3,17,0}$ & $-3.4$ & $120$ & $4.8 \pm 1.6$ \tabularnewline 
        \hline 
        $217.140$ & aGg’-(CH$_2$OH)$_2$ & $21_{4,17,0} \rightarrow 20_{4,16,1}$ & $-3.6$ & $124$ & $0.0193$ \tabularnewline 
        $217.450$ & aGg’-(CH$_2$OH)$_2$ & $24_{1,24,0} \rightarrow 23_{1,23,1}$ & $-3.6$ & $136$ & $0.0196$ \tabularnewline 
        $217.450$ & aGg’-(CH$_2$OH)$_2$ & $24_{0,24,0} \rightarrow 23_{0,23,1}$ & $-3.7$ & $136$ & $0.0152$ \tabularnewline 
        $217.588$ & aGg’-(CH$_2$OH)$_2$ & $21_{2,19,1} \rightarrow 20_{2,18,0}$ & $-3.7$ & $117$ & $0.0178$ \tabularnewline 
        $218.706$ & aGg’-(CH$_2$OH)$_2$ & $22_{12,11,0} \rightarrow 21_{12,10,1}$ & $-3.8$ & $195$ & $6.2 \times 10^{-3}$ \tabularnewline 
        $218.706$ & aGg’-(CH$_2$OH)$_2$ & $22_{12,10,0} \rightarrow 21_{12,9,1}$ & $-4.0$ & $195$ & $4.8 \times 10^{-3}$ \tabularnewline 
        $219.090$ & aGg’-(CH$_2$OH)$_2$ & $22_{10,13,0} \rightarrow 21_{10,12,1}$ & $-3.8$ & $174$ & $9.1 \times 10^{-3}$ \tabularnewline 
        $219.385$ & aGg’-(CH$_2$OH)$_2$ & $26_{11,16,1} \rightarrow 26_{10,16,0}$ & $-5.5$ & $230$ & $1.05 \times 10^{-4}$ \tabularnewline 
        $219.385$ & aGg’-(CH$_2$OH)$_2$ & $22_{9,14,0} \rightarrow 21_{9,13,1}$ & $-3.7$ & $164$ & $0.0107$ \tabularnewline 
        $219.385$ & aGg’-(CH$_2$OH)$_2$ & $26_{11,15,1} \rightarrow 26_{10,17,0}$ & $-5.4$ & $230$ & $1.35 \times 10^{-4}$ \tabularnewline 
        $219.385$ & aGg’-(CH$_2$OH)$_2$ & $22_{9,13,0} \rightarrow 21_{9,12,1}$ & $-3.8$ & $164$ & $8.3 \times 10^{-3}$ \tabularnewline 
        $219.386$ & aGg’-(CH$_2$OH)$_2$ & $45_{12,34,1} \rightarrow 44_{13,31,1}$ & $-5.4$ & $580$ & $5.3 \times 10^{-6}$ \tabularnewline 
        $219.581$ & aGg’-(CH$_2$OH)$_2$ & $22_{1,21,1} \rightarrow 21_{1,20,0}$ & $-3.6$ & $122$ & $0.021$ \tabularnewline 
        $219.765$ & aGg’-(CH$_2$OH)$_2$ & $20_{4,16,1} \rightarrow 19_{4,15,0}$ & $-3.6$ & $114$ & $0.021$ \tabularnewline 
        $220.497$ & aGg’-(CH$_2$OH)$_2$ & $22_{7,15,0} \rightarrow 21_{7,14,1}$ & $-3.8$ & $149$ & $0.0109$ \tabularnewline 
        \hline 
    \end{tabular} 
    \tablefoot{Frequency shows the rest frequency of the transitions in GHz; Transition shows the $J_{\rm K_a,K_c}$ where $J$ is the total angular momentum and $K_a$ and $K_c$ are the projections of J onto the $a$ and $c$ inertial axis; log $I$ is the logarithm of the line intensity in $\rm nm^2\,MHz$; $E_{\rm up}$ is the energy of higher level in K, and $\tau$ is the optical depth of the transitions.}
    \renewcommand*{\arraystretch}{1.0} 
    \label{table_transitions_maps}
    \end{table*}

\section{Zoomed-in line identification spectra}

\begin{figure*}[!h]
    \centering
    \includegraphics[width=\linewidth, valign=c]{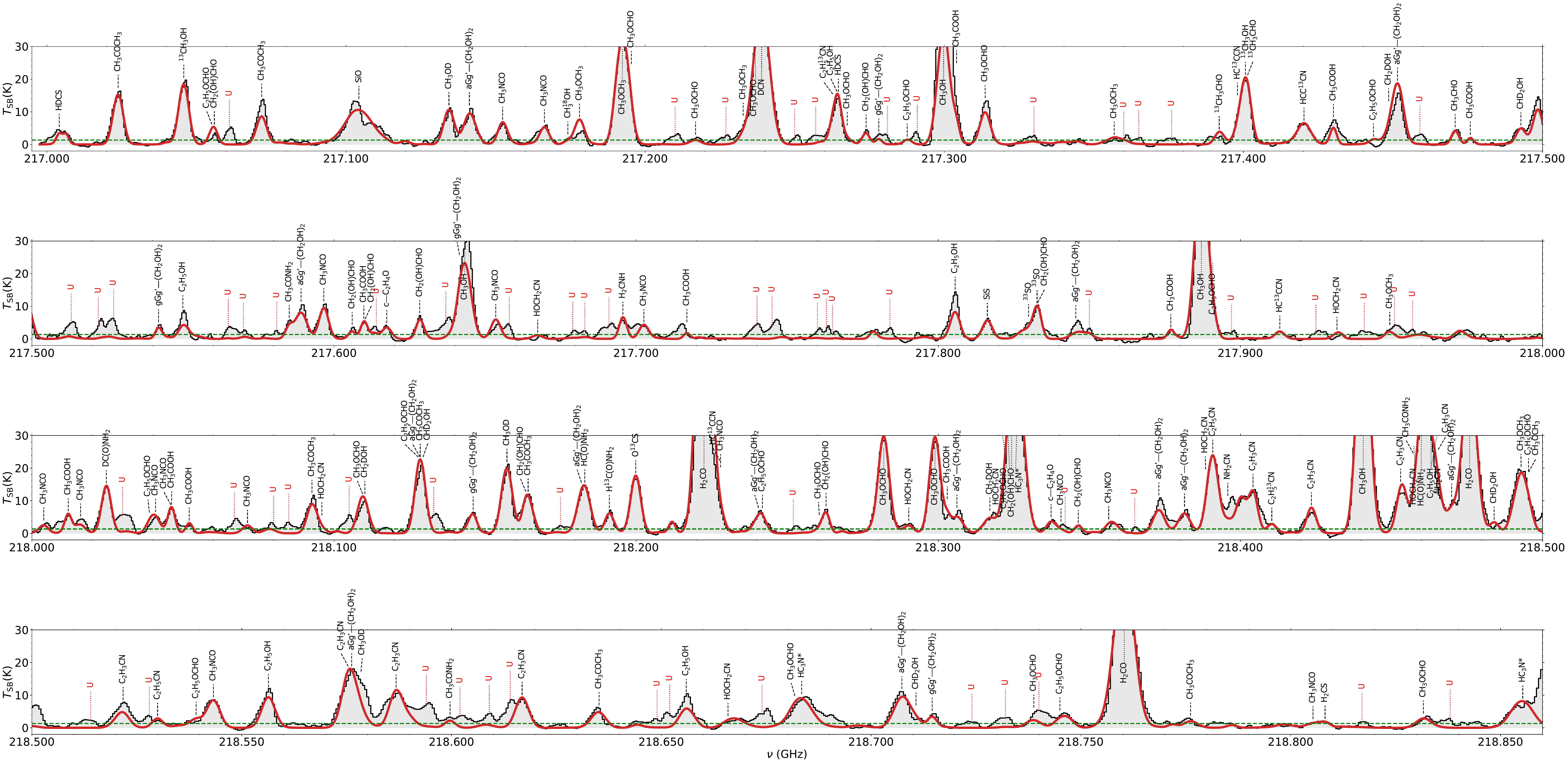}
    \caption{Same as Fig. \ref{fig:madcuba_spw0} but with a zoomed-in y-axis to better highlight the weaker transitions and the quality of the spectral fit. }
    \label{fig:madcuba_spw0_zoom}
\end{figure*}

\begin{figure*}[!h]
    \centering
     \includegraphics[width=\linewidth, valign=c]{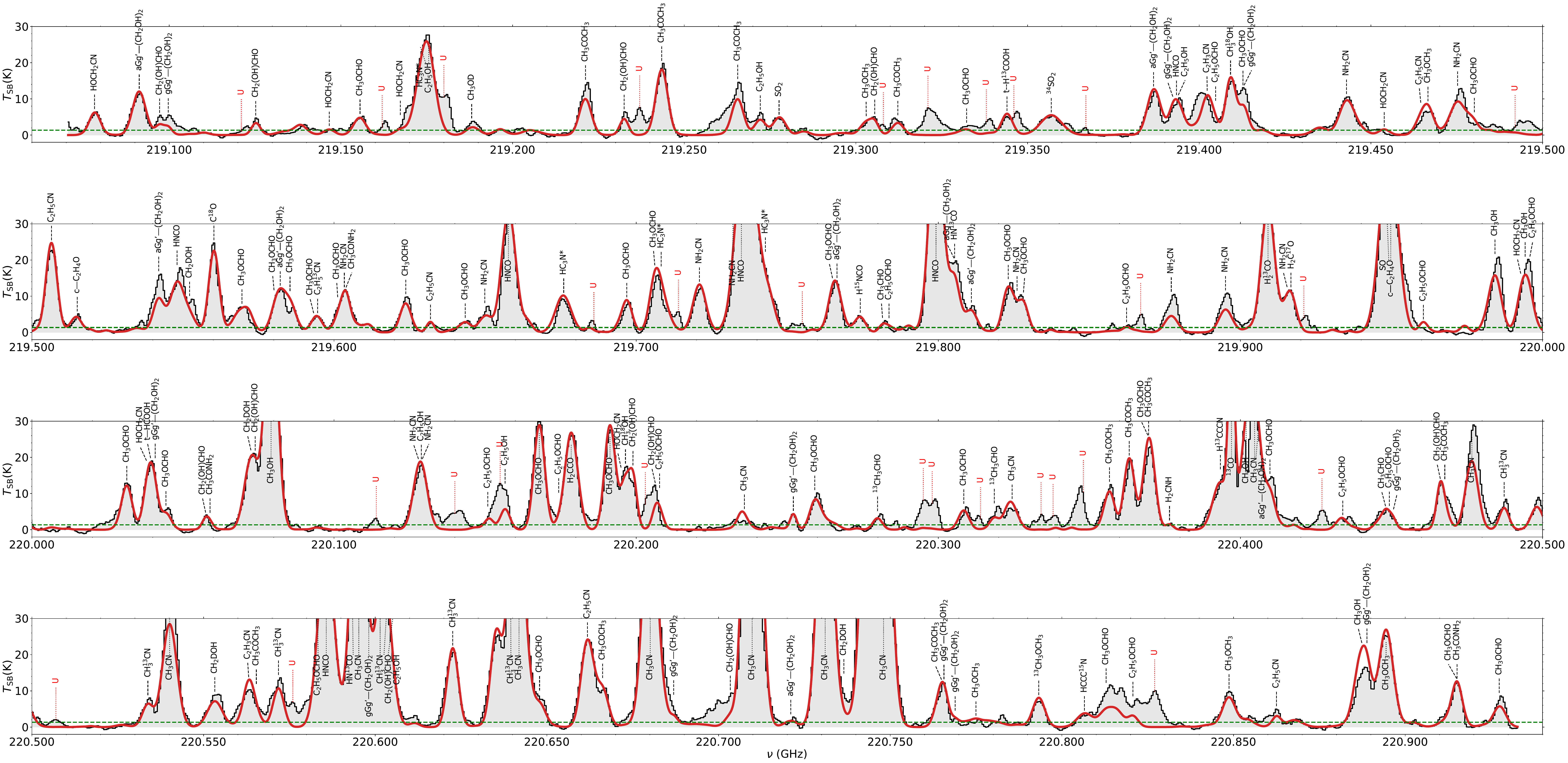}
    \caption{Same as Fig. \ref{fig:madcuba_spw1} but with a zoomed-in y-axis to better highlight the weaker transitions and the quality of the spectral fit.}
    \label{fig:madcuba_spw1_zoom}
\end{figure*}

Figures~\ref{fig:madcuba_spw0_zoom} and \ref{fig:madcuba_spw1_zoom} presents zoomed-in versions of the spw0 and spw1 spectra, respectively, shown in Sect.~\ref{sect:results}, aimed at better illustrating the weaker molecular lines used for line identification. 

\label{zoom spw}

\section{Complete version of physical parameters table}

Table~\ref{Tableapp:physical_parameters} presents the complete set of physical parameters derived from the SLIM analysis towards AG318$-$c9. Only a portion of this table is shown in the main text (Table~\ref{Table:physical_parameters}).

\renewcommand*{\arraystretch}{1.2}
\begin{table*}[!b] 
\centering
\caption{Results of the SLIM fits of the species analysed towards AG318$-$c9, sorted by increasing number of atoms.}
\begin{tabular}{@{\extracolsep{\fill}}ccccccc@{\extracolsep{\fill}}}
 \hline
{Species} & {$T_{\rm ex}$ (K)} & {$N$ ($\times 10^{16}$ cm$^{-2}$)} & {$X$ ($\times 10^{-9}$)} & {FWHM (km s$^{-1}$)} & {$V$ (km s$^{-1}$)} & {N$\degree$ transitions} \tabularnewline 
\hline 
$^{13}$CO$^{(a)}$ & $167$ & $42 \pm 3$ & $57 \pm 4$  & $3.2 \pm 0.3$ & $-31.59 \pm 0.12$ & 1\tabularnewline 
C$^{18}$O & $167$ & $30.4 \pm 1.3$ & $41.3 \pm 1.7$ & $5.4 \pm 0.3$ & $-33.66 \pm 0.11$ & 1\tabularnewline 
SO$^{(b)}$ & & $12.2 \pm 1.1$ & $16.5 \pm 1.4$ & $9.0 \pm 0.3$ & $-33.34 \pm 0.16$ & 1\tabularnewline 
$^{33}$SO & $167$ & $0.136 \pm 0.010$ & $0.185 \pm 0.014$ & $5.0$ & $-33.9 \pm 0.3$ & 4 \tabularnewline 
SiO & $167$ & $0.054 \pm 0.002$ & $0.073 \pm 0.003$ & $15.0 \pm 0.6$ & $-32.9 \pm 0.3$ & 1 \tabularnewline 
SiS & $167$ & $0.039 \pm 0.003$ & $0.053 \pm 0.004$ & $5.2 \pm 0.5$ & $-31.9 \pm 0.2$ & 4\tabularnewline 
SO$_2$$^{(b)}$ &  & $24.3 \pm 1.4$ & $33.1 \pm 1.9$ & $5.7 \pm 0.6$ & $-36.0 \pm 0.2$ & 2\tabularnewline 
$^{34}$SO$_2$ & $167$ & $0.77 \pm 0.07$ & $1.05 \pm 0.10$ & $10.0$ & $-36.4 \pm 0.5$ & 2\tabularnewline 
DCN & $167$ & $0.130 \pm 0.002$ & $0.178 \pm 0.003$ & $8.13 \pm 0.15$ & $-34.35 \pm 0.07$ & 1 \tabularnewline 
O$^{13}$CS & $167$ & $0.892 \pm 0.017$ & $1.21 \pm 0.02$ & $5.26 \pm 0.12$ & $-35.02 \pm 0.05$ & 1\tabularnewline 
H$_2$CO$^{(b)}$ &  & $33.9 \pm 7.3$ & $46.1 \pm 9.9$ & $7.0$ & $-34.14 \pm 0.06$ & 2\tabularnewline 
H$_2^{13}$CO & $167$ & $0.727 \pm 0.009$ & $0.989 \pm 0.012$ & $6.66 \pm 0.09$ & $-34.66 \pm 0.04$ & 2\tabularnewline 
H$_2$C$^{17}$O & $167$ & $0.125 \pm 0.011$ & $0.170 \pm 0.016$ & $6.0$ & $-36.4 \pm 0.3$ & 1\tabularnewline 
H$_2$CS & $167$ & $14.8 \pm 1.4$ & $20 \pm 2$ & $4.0 \pm 0.4$ & $-35.02 \pm 0.19$ & 1\tabularnewline 
HNCO$^{(b)}$ &  & $7.9 \pm 1.7$ & $10.8 \pm 2.3$ & $7.17 \pm 0.11$ & $-35.13 \pm 0.04$ & 10\tabularnewline 
H$^{15}$NCO & $167$ & $1.46 \pm 0.12$ & $1.98 \pm 0.16$ & $5.5 \pm 0.5$ & $-31.5 \pm 0.2$ & 1 \tabularnewline 
HN$^{13}$CO & $167$ & $0.17 \pm 0.03$ & $0.23 \pm 0.03$ & $5.0$ & $-35.1 \pm 0.5$ & 19\tabularnewline 
HDCS & $167$ & $0.13 \pm 0.02$ & $0.18 \pm 0.03$ & $3.0 \pm 0.6$ & $-35.1 \pm 0.3$ & 3\tabularnewline 
H$_2$CCO & $167$ & $1.56 \pm 0.04$ & $2.13 \pm 0.06$ & $6.4 \pm 0.2$ & $-35.07 \pm 0.09$ & 1\tabularnewline 
HC$_3$N*$^{(b)}$ &  & $0.84 \pm 0.18$ & $1.1 \pm 0.2$ & $8.08 \pm 0.17$ & $-35.04 \pm 0.07$ & 7\tabularnewline 
HC$^{13}$CCN & $167$ & $(6.5 \pm 1.0) \cdot 10^{-3}$ & $(8.9 \pm 1.4) \cdot 10^{-3}$ & $5.0$ & $-33.1 \pm 0.5$ & 1\tabularnewline 
HCC$^{13}$CN & $167$ & $0.0190 \pm 0.0013$ & $0.0258 \pm 0.0018$ & $7.1 \pm 0.6$ & $-34.8 \pm 0.2$ & 1\tabularnewline 
H$^{13}$CCCN & $167$ & $0.044 \pm 0.008$ & $0.060 \pm 0.010$ & $7.7 \pm 1.5$ & $-38.1 \pm 0.7$ & 1\tabularnewline
HCCC$^{15}$N & $167$ & $0.011 \pm 0.003$ & $0.015 \pm 0.004$ & $7.0$ & $-35$ & 1\tabularnewline 
HC$^{13}$CCN, v$_7=1$ & $167$ & $0.031 \pm 0.003$ & $0.042 \pm 0.003$ & $5.0$ & $-33.9 \pm 0.2$ & 1\tabularnewline 
NH$_2$CN & $202 \pm 10$ & $0.216 \pm 0.009$ & $0.294 \pm 0.013$ & $6.8 \pm 0.2$ & $-35.56 \pm 0.08$ & 19 \tabularnewline 

\hline 
\end{tabular}
\label{Tableapp:physical_parameters}
\end{table*}
\renewcommand*{\arraystretch}{1.0} 

\renewcommand*{\arraystretch}{1.2}
\begin{table*}[!t] 
\centering
\begin{tabular}{@{\extracolsep{\fill}}ccccccc@{\extracolsep{\fill}}}
\hline
{Species} & {$T_{\rm ex}$ (K)} & {$N$ ($\times 10^{16}$ cm$^{-2}$)} & {$X$ ($\times 10^{-9}$)} & {FWHM (km s$^{-1}$)} & {$V$ (km s$^{-1}$)} & N$\degree$ transitions \tabularnewline  \hline
H$_2$CNH & $250 \pm 80$ & $49 \pm 50$ & $67 \pm 70$ & $4.0$ & $-38.1 \pm 0.3$ & 3\tabularnewline 
t$-$HCOOH$^{(b)}$ & & $29.4 \pm 6.3$ & $40 \pm 8.6$ & $5.7 \pm 0.2$ & $-35.58 \pm 0.09$ & 1\tabularnewline 
t$-$H$^{13}$COOH & $167$ & $0.63 \pm 0.12$ & $0.86 \pm 0.16$ & $5.0$ & $-37$ & 1\tabularnewline  
CH$_3$OH$^{(b)}$ &  & $(9.4 \pm 2)\times 10^2$ & $(12.9 \pm 2.8)\times10^{2}$ & $6.61 \pm 0.10$ & $-34.60 \pm 0.04$ & 24\tabularnewline 
$^{13}$CH$_3$OH & $143 \pm 8$ & $20.5 \pm 0.5$ & $27.9 \pm 0.7$ & $5.37 \pm 0.16$ & $-35.35 \pm 0.07$ & 4\tabularnewline 
CH$_3^{18}$OH & $64 \pm 12$ & $2.2 \pm 0.2$ & $3.1 \pm 0.3$ & $4.4 \pm 0.3$ & $-35.59 \pm 0.15$ & 3\tabularnewline 
CH$_3$OD & $94 \pm 7$ & $11.7 \pm 1.3$ & $15.9 \pm 1.8$ & $5.6 \pm 0.4$ & $-35.91 \pm 0.16$ & 4\tabularnewline 
CH$_2$DOH & $130 \pm 7$ & $17.1 \pm 1.1$ & $23.3 \pm 1.6$ & $6.9 \pm 0.4$ & $-34.74 \pm 0.16$ & 6\tabularnewline 
CHD$_2$OH & $140 \pm 70$ & $4.3 \pm 1.9$ & $6 \pm 3$ & $5.0$ & $-36$ & 4\tabularnewline 
CH$_3$CN$^{(b)}$ & & $6.6 \pm 1.4$ & $9 \pm 1.9$ & $6.57 \pm 0.13$ & $-34.87 \pm 0.06$ & 6\tabularnewline 
CH$_3^{13}$CN & $127 \pm 19$ & $0.162 \pm 0.012$ & $0.221 \pm 0.017$ & $5.0$ & $-35.76 \pm 0.14$ & 6\tabularnewline 
HC(O)NH$_2$$^{(b)}$ & & $5.1 \pm 1.1$ & $6.9 \pm 1.5$ & $5.2 \pm 0.2$ & $-34.73 \pm 0.11$ & 5\tabularnewline 
H$^{13}$C(O)NH$_2$ & $167$ & $0.111 \pm 0.008$ & $0.150 \pm 0.010$ & $4.8 \pm 0.4$ & $-36.44 \pm 0.16$ & 2 \tabularnewline 
DC(O)NH$_2$ & $167$ & $0.30 \pm 0.02$ & $0.41 \pm 0.03$ & $5.0$ & $-37.3 \pm 0.2$ & 1\tabularnewline 
HC(O)NH$_2$, v$_{12}=1$ & $167$ & $3.43 \pm 0.14$ & $4.67 \pm 0.19$ & $6.3 \pm 0.3$ & $-35.59 \pm 0.12$ & 1\tabularnewline 
CH$_3$CHO$^{(b)}$ &  & $20.7 \pm 4.5$ & $28.2 \pm 6.1$ & $4.0$ & $-35.90 \pm 0.16$ & 7\tabularnewline 
$^{13}$CH$_3$CHO & $180 \pm 90$ & $0.4 \pm 0.4$ & $0.6 \pm 0.6$ & $5.0$ & $-35.5 \pm 0.4$ & 6\tabularnewline 
$\rm c-C_2H_4O$ & $128 \pm 11$ & $0.71 \pm 0.05$ & $0.97 \pm 0.07$ & $5.0 \pm 0.4$ & $-36.14 \pm 0.16$ & 5 \tabularnewline 
CH$_3$NCO & $92 \pm 8$ & $0.82 \pm 0.06$ & $1.11 \pm 0.08$ & $5.0$ & $-35.80 \pm 0.09$ & 13 \tabularnewline 
HOCH$_2$CN & $51 \pm 3$ & $1.28 \pm 0.09$ & $1.75 \pm 0.13$ & $5.0$ & $-37.04 \pm 0.18$ & 6\tabularnewline 
C$_2$H$_3$CN & $138 \pm 14$ & $0.360 \pm 0.012$ & $0.489 \pm 0.017$ & $5.0$ & $-36.17 \pm 0.11$ & 18\tabularnewline 
CH$_3$OCHO & $167 \pm 3$ & $22.0 \pm 0.4$ & $29.9 \pm 0.6$ & $5.59 \pm 0.08$ & $-35.22 \pm 0.04$ & 41\tabularnewline 
CH$_2$(OH)CHO & $168 \pm 13$ & $1.45 \pm 0.12$ & $1.97 \pm 0.17$ & $4.0$ & $-36.89 \pm 0.10$ & 19\tabularnewline 
CH$_3$COOH & $70 \pm 30$ & $0.74 \pm 0.07$ & $1.00 \pm 0.10$ & $3.5$ & $-36.54 \pm 0.17$ & 9\tabularnewline 
CH$_3$COOH, vt=1 & $167$ & $2.38 \pm 0.18$ & $3.2 \pm 0.2$ & $3.5$ & $-36$ & 24\tabularnewline 
CH$_3$OCH$_3$$^{(b)}$ &  & $85.6 \pm 18.4$ & $116.5 \pm 25$ & $5.16 \pm 0.17$ & $-35.19 \pm 0.06$ & 21\tabularnewline 
$^{13}$CH$_3$OCH$_3$ & $133 \pm 16$ & $1.81 \pm 0.19$ & $2.5 \pm 0.3$ & $5.5 \pm 0.4$ & $-35.10 \pm 0.15$ & 19\tabularnewline 
C$_2$H$_5$OH & $149 \pm 14$ & $10.2 \pm 0.8$ & $13.8 \pm 1.1$ & $5.1 \pm 0.2$ & $-36.32 \pm 0.10$ & 10\tabularnewline 
C$_2$H$_5$CN$^{(b)}$ & & $5.3 \pm 1.1$ & $7.2 \pm 1.5$ & $5.6$ & $-35.02 \pm 0.05$ & 8\tabularnewline 
C$_2$H$_5^{13}$CN & $167$ & $0.117 \pm 0.014$ & $0.159 \pm 0.019$ & $5.0$ & $-36$ & 5\tabularnewline 
C$_2$H$_5$CN, v$_{20}=1-\rm A$ & $167$ & $2.4 \pm 0.3$ & $3.2 \pm 0.4$ & $4.0$ & $-33$ & 2\tabularnewline 
CH$_3$CONH$_2$ & $70 \pm 20$ & $0.122 \pm 0.014$ & $0.166 \pm 0.019$ & $4.0$ & $-37.1 \pm 0.3$ & 25\tabularnewline 
$\rm aGg^{\prime}-(CH_2OH)_2$ & $240 \pm 30$ & $5.1 \pm 0.6$ & $6.9 \pm 0.9$ & $6.19 \pm 0.19$ & $-35.90 \pm 0.08$ & 37 \tabularnewline 
$\rm gGg^{\prime}-(CH_2OH)_2$ & $167$ & $1.5 \pm 0.2$ & $2.1 \pm 0.3$ & $3.5$ & $-36.9 \pm 0.3$ & 9 \tabularnewline 
CH$_3$COCH$_3$ & $84 \pm 6$ & $1.4 \pm 0.6$ & $1.9 \pm 0.8$ & $5.33 \pm 0.11$ & $-35.59 \pm 0.05$ & 39 \tabularnewline 
$\rm C_2H_5OCHO$ & $110 \pm 30$ & $1.5 \pm 0.2$ & $2.1 \pm 0.3$ & $4.5$ & $-37$ & 38\tabularnewline 
\hline 
\end{tabular}
\tablefoot{
 The derived physical parameters ($T_{\rm ex}$, $N$, $X=N_{\rm species}/N_{\rm H_2}$, FWHM, and $V$) are listed along with their associated uncertainties. The last column indicates the number of transitions used to fit the spectrum. The parameters we fixed during the fitting procedure are reported without uncertainties.
When $T_{\rm ex}$ is fixed, $T_{\rm ex}^{\rm CH_3OCHO}=167\,\rm K$ is assumed (see Sect.\,\ref{macro}). HC$_3$N* represents the group composed of
HC$_3$N, HC$_3$N ($v_6=1$), HC$_3$N ($v_7=1$), and HC$_3$N ($v_7=2$).
\tablefoottext{a}{Species exhibiting a unique self-absorbed transition,
resulting in a lower limit on the column density.}
\tablefoottext{b}{The column density was estimated from the less abundant isotopologues using the isotopic ratios $\rm ^{12}C/^{13}C = 46.6\pm10$ \citep{Yan_2019}, $\rm ^{32}S/^{34}S = 20.7\pm1.2$ \citep{yan2023}, and $\rm ^{32}S/^{33}S = 88.9\pm7.7$ \citep{yan2023}.}}
\end{table*}
\renewcommand*{\arraystretch}{1.0}
\end{appendix}
\end{document}